\documentclass[twocolumn,superscriptaddress,amsmath,amssymb,aps,nofootinbib,reprint]{revtex4-2}

\usepackage{wasysym}
\usepackage{tikz}
\usepackage[unicode=true,bookmarks=true,bookmarksnumbered=false,bookmarksopen=false,breaklinks=false,backref=false,colorlinks=true]{hyperref}
\hypersetup{urlcolor=blue,linkcolor=violet,citecolor=blue}
\usepackage[nameinlink,capitalize]{cleveref}
\usepackage{natbib}
\usepackage{dcolumn}
\usepackage{bm}
\usepackage{oplotsymbl}
\usepackage{graphicx}
\graphicspath{ {./images/} }
\usepackage{physics}
\usepackage[english]{babel}
\usepackage{xcolor}
\usepackage[normalem]{ulem}
\usepackage{mathtools}
\usepackage{tabularx}
\usepackage{array}
\usepackage{booktabs}
\usepackage{sansmath}
\usepackage{bm}
\usepackage{yhmath}
\usepackage{orcidlink}
\usepackage[normalem]{ulem}
\usepackage[caption=false]{subfig}
\usepackage{ragged2e}
\usepackage{enumitem}
\usepackage{adjustbox}
\usepackage{svg}
\usepackage{float}
\crefname{section}{Sec.}{Secs.}

\definecolor{col_dim}{RGB}{53,132,228}
\definecolor{col_mon}{RGB}{255,163,72}
\definecolor{col_link}{RGB}{119,118,123}
\definecolor{col_membrane}{RGB}{38,162,105}

\newcommand\scl{0.15}
\newcommand{\monomerhopA}{\hspace{0.5mm}\begin{tikzpicture}[scale=\scl]
 \draw[line width=0.4mm, col_link] (0,0) -- (1,0);
 \draw[line width=0.25mm, pink] (1,0) -- ++(72:1);
 \filldraw[col_dim] (0,0) circle (6pt);
  \filldraw[col_dim] (1,0) circle (6pt);
    \filldraw[col_mon] (1,0) ++(72:1) circle (6pt);
 \end{tikzpicture} \hspace{0.5mm}  }
 
\newcommand{\monomerhopB}{\hspace{0.5mm}\begin{tikzpicture}[scale=\scl]
 \draw[line width=0.20mm, pink] (0,0) -- (1,0);
 \draw[line width=0.4mm, col_link] (1,0) -- ++(72:1);
 \filldraw[col_mon] (0,0) circle (6pt);
  \filldraw[col_dim] (1,0) circle (6pt);
    \filldraw[col_dim] (1,0) ++(72:1) circle (6pt);
 \end{tikzpicture} \hspace{0.5mm}  } 

\newcommand{\plaqA}{ \hspace{0.5mm} \begin{tikzpicture}[scale=\scl]
 \draw[line width=0.20mm, pink] (0,0) -- (1,0);
\draw[line width=0.4mm, col_link] (1,0) -- ++(72:1);
  \draw[line width=0.25mm, pink] (1,0)++(72:1) -- (72:1);
  \draw[line width=0.4mm, col_link] (0,0) -- ++(72:1);
 \filldraw[col_dim] (0,0) circle (6pt);
  \filldraw[col_dim] (1,0) circle (6pt);
    \filldraw[col_dim] (1,0) ++(72:1) circle (6pt);
    \filldraw[col_dim] (72:1) circle (6pt);
 \end{tikzpicture} \hspace{0.5mm}}

 \newcommand{\plaqB}{ \hspace{0.5mm} \begin{tikzpicture}[scale=\scl]
 \draw[line width=0.40mm, col_link] (0,0) -- (1,0);
\draw[line width=0.25mm, pink] (1,0) -- ++(72:1);
  \draw[line width=0.4mm, col_link] (1,0)++(72:1) -- (72:1);
  \draw[line width=0.25mm, pink] (0,0) -- ++(72:1);
 \filldraw[col_dim] (0,0) circle (6pt);
  \filldraw[col_dim] (1,0) circle (6pt);
    \filldraw[col_dim] (1,0) ++(72:1) circle (6pt);
    \filldraw[col_dim] (72:1) circle (6pt);
 \end{tikzpicture} \hspace{0.5mm}}

\begin{document}

\title{A quantum monomer-dimer model on Penrose tilings}

\author{Jeet Shah\,\orcidlink{0000-0001-5873-8129}}
\affiliation{Joint Quantum Institute, NIST/University of Maryland, College Park, Maryland 20742, USA}
\affiliation{Joint Center for Quantum Information and Computer Science, NIST/University of Maryland, College Park, Maryland 20742, USA}
\author{Gautam Nambiar\,\orcidlink{0000-0003-4305-8600}}
\affiliation{Joint Quantum Institute, NIST/University of Maryland, College Park, Maryland 20742, USA}
\author{Alexey~V.~Gorshkov\,\orcidlink{0000-0003-0509-3421}}
\affiliation{Joint Quantum Institute, NIST/University of Maryland, College Park, Maryland 20742, USA}
\affiliation{Joint Center for Quantum Information and Computer Science, NIST/University of Maryland, College Park, Maryland 20742, USA}
\author{Victor Galitski}
\affiliation{Joint Quantum Institute, NIST/University of Maryland, College Park, Maryland 20742, USA}

\begin{abstract}

We define a quantum monomer-dimer model in the space of maximal dimer coverings of quasicrystalline Penrose tilings.
Since Penrose tilings do not admit perfect dimer coverings, as shown by F.~Flicker et al., PRX~\textbf{10},~011005~(2020), monomers are necessarily present in our model.
The model features a frustration-free Rokhsar-Kivelson (RK) point where the ground state is a uniform superposition of all the exponentially many maximal dimer coverings, despite the presence of a finite density of monomers.
We map our model to a $\mathbb{Z}_2$ gauge theory with matter and calculate various correlators to characterize the phase of the system at the RK point using classical Monte Carlo calculations. 
Specifically, we compute the dimer-dimer and vison-vison correlators, as well as  open Wilson lines and closed Wilson loops corresponding to the monomers and the visons. 
We find that both the dimer-dimer and vison-vison correlators decay exponentially with distance.
 The open Wilson lines and closed Wilson loops decay exponentially with the same correlation length, indicating that the gauge theory is in the confined phase, which implies that the system is likely in an ordered phase.
  
\end{abstract}

\maketitle

\textit{Introduction.---}Quantum dimer models are constrained models where the fundamental degrees of freedom called dimers occupy two neighboring sites of a lattice. 
In 1988, Rokhsar and Kivelson (RK) proposed the first quantum dimer model~\cite{rokhsar1988superconductivity}.
Since then quantum dimer models have been extensively studied on a number of lattices in one, two, and three dimensions~\cite{natalia2019dmrg,balents2002fractionalization, moessner2003threedimensional, moessner2001resonating, hermele2004pyrochlore, savary2017quantum, balasubramanian2022classical, balasubramanian2024interplay,shannon2012quantum, sikora2009quantum,misguich2002quantum, yan2022triangular,moessner2001phase,moessner2011introduction,shah2025quantum,senthil2004deconfined}.
Depending on the lattice and the dimensionality, quantum dimer models exhibit a rich set of phases possessing exotic properties.
The original quantum dimer model proposed by Rokhsar and Kivelson on the square lattice was later shown to have an ordered ground state with dimer defects (monomers) being linearly confined (i.e.~their interaction potential grows linearly with distance)~\cite{yan2021widely,syljuasen2006plaquette,syljuasen2005continuous,ralko2008generic,banerjee2014interfaces}. 
The situation is more interesting in the triangular-lattice dimer model, where the ground state was found to possess $\mathbb{Z}_2$ topological order with fractionalized excitations~\cite{moessner2001resonating,moessner2001rvb,misguich2008quantum,ioselevich2002groundstate} and nontrivial topological entanglement entropy in an \textit{extended} region of the phase diagram.
In three dimensions, a quantum dimer model on the frustrated pyrochlore lattice  hosts a $U(1)$ quantum spin liquid with fractionalized electric charges, loop-like magnetic monopoles, and gapless photons~\cite{hermele2004pyrochlore,sikora2009quantum,shannon2012quantum,pace2021emergent,shah2025quantum}. 
These examples show that quantum dimer models have very distinct behavior depending on the lattice on which they are defined.
\begin{figure}[t]
\captionsetup[subfigure]{labelformat=empty,captionskip=-20pt}
 \centering
 \hfill
 \subfloat[\label{fig:maximal-covering}]{%
  \includegraphics[width=0.49\columnwidth]{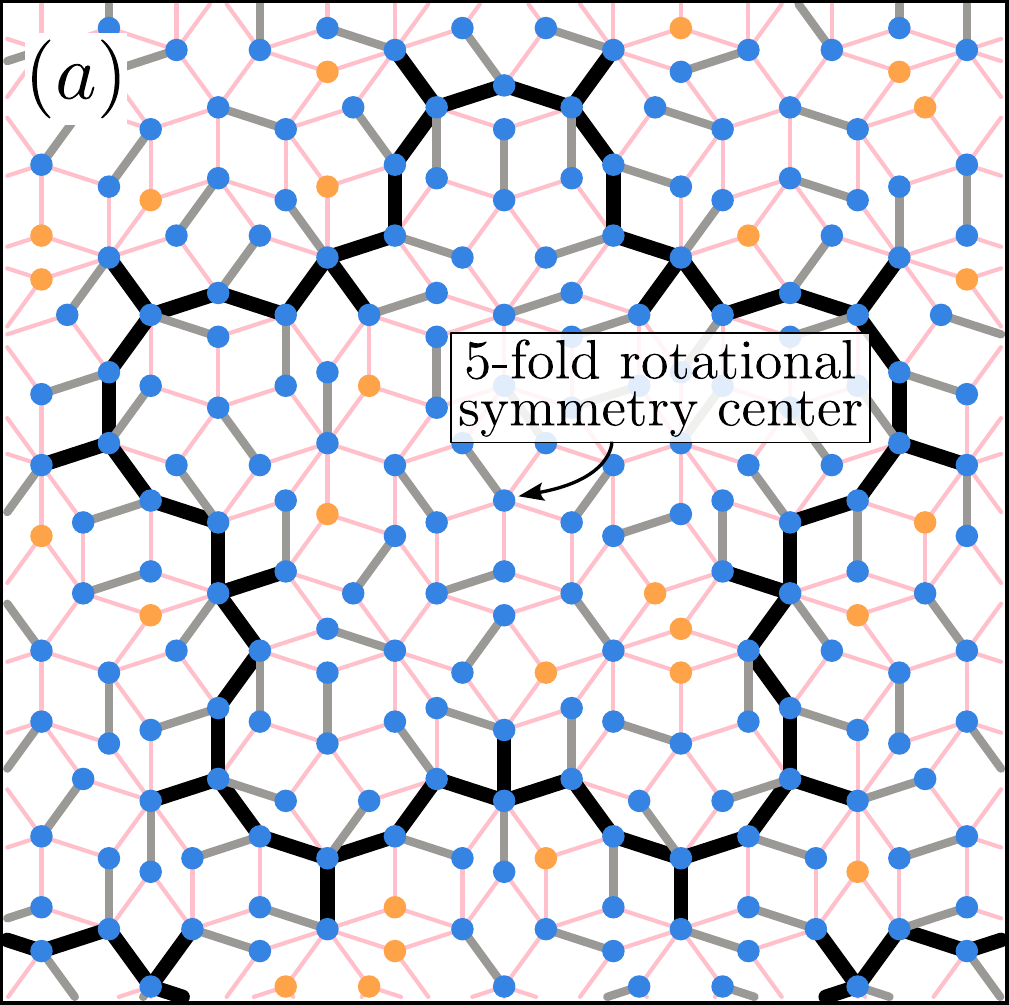}%
}
     \hfill
     \subfloat[\label{fig:fm-loops}]{%
     \includegraphics[width=0.49\columnwidth]{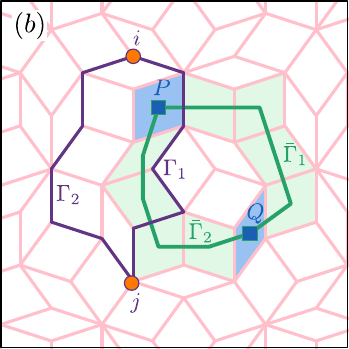}%
     }
    \caption{(a) A maximal dimer covering of a Penrose tiling.
    Monomers are shown by the orange dots, dimers by gray links with blue dots at their ends, and links of monomer membranes by thick black lines. (b) Example of paths for the monomer and vison Fredenhagen-Marcu (FM) order parameters. $\Gamma_{1,2}$ are two paths for the monomer FM order parameter between sites $i$ and $j$, shown in orange. $\hat{M} (\Gamma_{i})$ is the operator that moves monomers between sites $i$ and $j$ along $\Gamma_{i}$ where $i \in \{1, 2\}$. $\overline{\Gamma}_{i}$ are two valid vison paths, i.e., they intersect an even number of links of all plaquettes except $P$ and $Q$. $\hat{V} (\overline{\Gamma}_i)$ are  operators that create two visons on the blue plaquettes $P$ and $Q$.  }
\end{figure}
It is only in the last few years that attention has gone towards \textit{classical} dimer models on more complex quasicrystalline lattices~\cite{flicker2020classical,lloyd2022statistical, singh2024exact}, which can be viewed as projections of periodic higher-dimensional lattices onto lower-dimensional manifolds~\cite{debruijn1981algebraic,debruijn1981algebraic2}. 
Periodic tilings are translationally invariant and can only have two-fold, three-fold, four-fold, and six-fold rotational symmetries according to the crystallographic restriction theorem. 
However, aperiodic tilings, which include quasicrystals~\cite{kalugin1986electron,kalugin1985almn,fujiwara1991universal,repetowicz1998exact,collins2017imaging,levine1986quasicrystals}, can have other rotational symmetries such as five-fold (Penrose tiling~\cite{penrose1974therole}) and eight-fold (Ammann–Beenker tiling~\cite{beenker1982algebraic}). 
In Ref.~\cite{flicker2020classical}, Flicker et al.~studied the dimer coverings of Penrose tilings and proved that a particular kind of Penrose tilings made of rhombuses, known as P3 Penrose tilings have a finite density of monomers in all maximal dimer coverings. 
Moreover, certain well-defined links of the tiling form loops and are never occupied by a dimer in any of the maximal dimer coverings. 
These loops are known as ``monomer membranes"~\cite{flicker2020classical}.
Thus, the dimer coverings of regions within these loops are independent of each other.
In \cref{fig:maximal-covering}, we show a maximal dimer covering of a P3 Penrose tiling.
The thick black links in this figure constitute a monomer membrane.
These interesting features of  dimer coverings of Penrose tilings, which are qualitatively different from those of regular lattices, motivate us to study a \textit{quantum} monomer-dimer model on  Penrose tilings. 
We point out that quantum and classical dimer models  with finite density of monomers have been studied previously on periodic lattices~\cite{syljuasen2005continuous, ralko2007phase, poilblanc2006doping, poilblanc2008properties, castelnovo2007zero,sreejith2014critical}.

In this Letter, we study a quantum monomer-dimer model on P3 Penrose tilings (henceforth called Penrose tilings). 
Since there are no perfect dimer coverings of Penrose tilings~\cite{flicker2020classical}, the inclusion of monomers (sites not touched by any dimer) is necessary.
Our Hamiltonian has an RK point where the exact ground state wavefunction can be determined, similar to Hamiltonians with monomers studied in Ref.~\cite{papanikolaou2007quantum,ribeiro2007single}
We map our model to a $\mathbb{Z}_2$ gauge theory with matter (monomers) and calculate various correlators in the RK wavefunction using a classical Metropolis Monte Carlo calculation.
The quantities we calculate are the dimer-dimer correlator and the vison-vison correlator, as well as open Wilson lines and closed Wilson loops corresponding to monomers and visons ~\cite{fredenhagen1983charged,fredenhagen1986,fredenhagen1988dual,marcu1986uses,bricmont1983order,gregor2011diagnosing,verresen2021,semeghini2021probing}.
We average these quantities over different pairs of endpoints or paths depending on the correlator.

We consider the following Hamiltonian defined in the space of maximal dimer coverings of Penrose tilings:
\begin{equation}
\begin{aligned}
    \label{eq:dimer-monomer-H}
    \hat{H} =&  \sum -t \left[ \ket{\monomerhopA}\bra{\monomerhopB} + \text{H.c.} \right] + U_{\text{RK}} \left[ \ket{\monomerhopA}\bra{\monomerhopA} + \ket{\monomerhopB}\bra{\monomerhopB} \right] \\
     \sum & -J \left[\ket{\plaqA} \bra{\plaqB} + \text{H.c.}\right] + V_{\text{RK}} \left[\ket{\plaqA} \bra{\plaqA} + \ket{\plaqB} \bra{\plaqB}\right],
\end{aligned}
\end{equation}
where orange dots denote monomers, and gray links with blue dots at their ends denote dimers.
The first sum is over all the pairs of links that share a vertex, while the second sum is over all the plaquettes [both types of rhombuses, see \cref{fig:maximal-covering}]. 
The first term ($\propto t$) gives rise to monomer hopping (a dimer simultaneously hops to keep the state in the maximal dimer covering sector). 
The third term ($\propto J$) is a ring exchange/resonance term for flippable plaquettes (plaquettes with dimers on two of their opposite sides). 
The second term ($\propto U_\text{RK}$) is an RK potential term for monomers, and the fourth term ($\propto V_\text{RK}$) is an RK potential for  dimers~\cite{rokhsar1988superconductivity}.
We also propose a microscopic spin Hamiltonian in the supplemental material (SM)~\cite{supp}, whose low energy limit is a quantum monomer-dimer model on Penrose tilings, having some of the terms of \cref{eq:dimer-monomer-H}.

We choose to study this Hamiltonian because, when $t = U_{\text{RK}}$ and $J = V_{\text{RK}}$, the Hamiltonian becomes a sum of projectors, and the ground state is annihilated by all the projectors:
\begin{equation}
    \label{eq:RK-hamiltonian}
    \begin{aligned}
        \hat{H}_\text{RK} = & t \sum \Big[ \ket{\monomerhopA} - \ket{\monomerhopB} \Big] \Big[ \bra{\monomerhopA} - \bra{\monomerhopB} \Big] \\
        & + J \sum \Big[ \ket{\plaqA} - \ket{\plaqB} \Big] \Big[ \bra{\plaqA } - \bra{\plaqB} \Big]~.
    \end{aligned}
\end{equation}
It can be easily verified that $\ket{\psi_{\text{RK}}} = \sum_{\mathcal{D}} \ket{\mathcal{D}}$, which is a uniform superposition of all maximal dimer coverings $\mathcal{D}$, is 
the ground state for any $t,J > 0$.
It was shown in Ref.~\cite{flicker2020classical} that monomer hoppings connect all maximal dimer coverings, i.e., starting from any maximal dimer covering, repeated monomer hoppings generate all the maximal dimer coverings. 
This implies that the ground state is unique and preserves the five-fold rotational symmetry.
The connectivity of the Penrose tilings constrains the maximal dimer coverings in a way that the links of monomer membranes, which form closed loops, do not host a dimer, and the sites of the monomer membranes do not host monomers in any of the maximal dimer coverings~\cite{flicker2020classical}.
These loops are called monomer membranes because a monomer that is inside one of these loops can never go outside of it by repeated monomer hops~\cite{flicker2020classical}.
Thus, one can remove the links of the monomer membranes from the model without affecting maximal dimer coverings.
Doing so separates the tiling into disconnected pieces which we label by $R_k$, where $k$ is an index that labels different disconnected pieces. Two such regions are shown in \cref{supp-fig:regions} of the SM~\cite{supp}.

Since $R_k$ are disconnected from each other, their dimer coverings are independent of each other, and the RK wavefunction $\ket{\psi_{\text{RK}}}$ separates into a tensor product of RK wavefunctions of the regions $R_k$.
That is, $\ket{\psi_{\text{RK}}} = \underset{k}{\otimes} \left[ \sum_{\mathcal{D}_{R_k}} \ket{\mathcal{D}_{R_k}} \right]$~\cite{normalization-footnote},
where the sum is over all the maximal dimer coverings $\mathcal{D}_{R_k}$ of a disconnected region $R_k$. 
This implies that there are no (connected) correlations between different regions $R_k$.
Schematically, some of the monomer membranes form concentric loops around the five-fold rotationally symmetric center of the tilings.
This can be seen from \cref{supp-fig:regions} of the SM~\cite{supp} and Fig.~7 of Ref.~\cite{flicker2020classical}.
Although monomer membranes might intuitively suggest confinement, this is not necessarily the case. 
A dimer Hamiltonian on a modified triangular lattice can be constructed where monomers are restricted to certain (infinite) regions, yet the monomers are deconfined.
Thus, it is not a priori clear whether our model is confining or deconfined.
Thus, by going far enough from the center, one can obtain a region $R_k$ that is as large as desired.
For these regions, we can meaningfully ask: what is the \textit{long}-distance behavior of the correlators that determines the phase?
To identify which correlators should be calculated to diagnose the phase of the system, we first map the monomer-dimer model to a $\mathbb{Z}_2$ gauge theory with matter.


\textit{Mapping to a $\mathbb{Z}_2$ gauge theory with matter.---}
Define $\mathbb{Z}_2$ gauge field variables $\hat{\sigma}_{ij}$ on the links of the Penrose tiling, where $i$ and $j$ are neighboring sites. 
If the link $(i,j)$ is occupied by a dimer, then $\sigma^z_{ij} = -1$, and if not, then $\sigma^z_{ij} = 1$. 
Next, define matter field variables $\hat{\tau}_i$ on the sites of the tiling such that $\tau_i^z = -1$ if a monomer is present, and $\tau_i^z = 1$ if a monomer is not present on site $i$.
The constraint that a site either has a monomer or is touched by a single dimer leads to the following Gauss's law for all dimer coverings $\ket{\mathcal{D}}$~\cite{gauss-footnote}:
\begin{equation}
    \label{eq:gauss-law}
    \hat{G}_i \ket{\mathcal{D}} = -\ket{\mathcal{D}}, \; \text{with}\; \hat{G}_i = \hat{\tau}_i^z \prod_{j \in \text{nbr}(i)} \hat{\sigma}_{ij}^z~, 
\end{equation}
where the product in $\hat{G}_i$ is over all neighbors $j$ of $i$.
Note that Gauss’s law alone does not fully enforce the hard-core dimer constraint, so we further restrict gauge field configurations to those obtained from valid maximal dimer coverings. 
The negative sign of the eigenvalue of $\hat G_i$ implies that the divergence of the $\mathbb{Z}_2$ electric field is nonzero indicating the presence of a background $\mathbb{Z}_2$ gauge charge on every site.
A monomer on site $i$ cancels the background charge at $i$, causing all electric field variables emanating from $i$ to take the value of $+1$ and the $\mathbb{Z}_2$ electric field is divergence-less.
Such a gauge theory is referred to as an odd $\mathbb{Z}_2$ gauge theory~\cite{moessner2001shortranged}. 
The embedding of the monomer-dimer model in a $\mathbb{Z}_2$ gauge theory with matter~\cite{fradkin1979phase,gregor2011diagnosing} allows one to identify the vison operator~\cite{read1989statistics,senthil2000z2gauge,ioselevich2002groundstate}.
The vison operator on a plaquette $P$ of a Penrose tiling is an operator that changes the $\mathbb{Z}_2$ gauge flux given by $\hat{\Phi}_P = \prod_{(i,j) \in \partial P} \hat{\sigma}_{ij}^x$ through the plaquette $P$, where $\partial P$ is the set of edges of $P$.
Since the flux is associated with plaquette centers, the visons live on the dual graph of the tiling. 
A single $\hat{\sigma}^z_{ij}$ operator changes the flux through two plaquettes that share the link $(i,j)$ and thus creates a vison on each of these plaquettes.
These visons can be separated onto plaquettes $P$ and $Q$ [see \cref{fig:fm-loops}] by applying a product of $\hat{\sigma}_{ij}^z$ along a path on the dual graph that goes from $P$ to $Q$.
These paths intersect two links of all plaquettes along the path except $P$ and $Q$ which are intersected only once.
Two such paths on the dual graph, $\overline{\Gamma}_1$ and $\overline{\Gamma}_2$, are shown in \cref{fig:fm-loops}.
Thus, the vison operator is given by 
\begin{equation}
    \hat{V} (\overline{\Gamma})   = \prod_{(i,j) \in \overline{\Gamma}} \hat{\sigma}^z_{ij} ~,
\end{equation} 
where $\overline{\Gamma}$ is a path on the dual graph.
Note that, due to the presence of monomers, the vison operator depends on the path $\overline{\Gamma}$ and not solely on its endpoints $P$ and $Q$. 
If the monomers were fixed by additional terms in the Hamiltonian, then the vison correlator $|\langle \hat{V} (\overline{\Gamma}) \rangle|$~\cite{vison-absolute-footnote} would have been the same for all paths that have the same endpoints.

Similarly, the operator that creates a monomer at site $i$ and destroys one at site $j$, or vice versa, is given by 
\begin{equation}\label{eq:monomer-op}
\hat{M}(\Gamma) = \hat{\mathcal{P}} \hat{\tau}_i^x \hat{\tau}_j^x \prod_{(k,l) \in \Gamma} \hat{\sigma}^x_{kl}~,
\end{equation}
where $\Gamma$ is a path from site $i$ to site $j$ along the edges of the Penrose tiling and $\hat{\mathcal{P}}$ is a projector onto the maximal dimer covering sector. 
\cref{fig:fm-loops} shows two example paths $\Gamma_1$ and $\Gamma_2$.
Note that, in a maximal dimer covering, it is possible to move a monomer from $i$ to $j$ along a path $\Gamma$ only if the links of $\Gamma$  alternate between dimers and non-dimers.
This operator also depends on the entire path $\Gamma$ and not solely on its endpoints. 
Unlike the vison operator, the operator in \cref{eq:monomer-op} would be path dependent even if the monomers were fixed by additional terms.
If a vison path $\overline{\Gamma}$ intersects a monomer path $\Gamma$, then $\hat{V}(\overline{\Gamma})$ anti-commutes with $\hat{M}(\Gamma)$, showing that  visons and monomers have a mutual statistical phase of $-1$.
In the gauge theory picture, the monomers (visons) are the electric (magnetic) charges.


\textit{Correlators.---}To characterize the phase of the gauge theory, we calculated the following quantities: the dimer-dimer connected correlator and the vison-vison correlator, as well as  open Wilson lines and closed Wilson loops corresponding to both visons and monomers. 
These were calculated for the RK wavefunction $\ket{\psi_{\text{RK}}}$ using classical Metropolis Monte Carlo for two regions $R_1$ and $R_2$ of a particular Penrose tiling which we show in~\cref{supp-fig:regions} of the SM~\cite{supp}.
We use the Hopcroft-Karp~\cite{hopcroft1973an} algorithm to generate the initial maximal dimer covering for these calculations.
We now  present our results.

\begin{figure}[htbp]
  \centering
  \includegraphics[width=0.99\columnwidth]{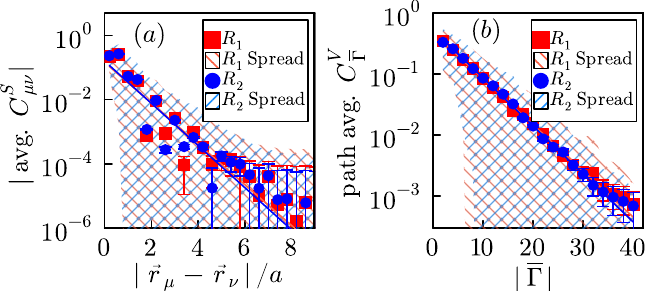}
  \caption{(a) Pairs of links $\mu, \nu \in R_{1,2}$ are binned together according to the Euclidean distance $\abs{\vec{r}_\mu-\vec{r}_\nu}$ between midpoints of the links, with a bin-width of $0.4a$. Here $a$ is the side length of the rhombuses. Bin-averaged $C^S_{\mu \nu}$ as a function of $\abs{\vec{r}_\mu-\vec{r}_\nu}/a$ is plotted in this Figure. 
  The error bars indicate the standard error of the averaged correlator, arising from a finite number of Monte Carlo samples.
  The hatched region indicates the standard deviation of the bins representing the spread of the correlator.
  We indicate the Monte Carlo error in the averaged correlator with error bars and the standard deviation of the bins with hatched regions in all plots of this paper.
  For a discussion of the binning procedure, see \cref{sec:dimer-dimer-averaging} of SM~\cite{supp}.
  The solid lines show exponential fits to the correlator in this and all subsequent plots.
  (b) Path-averaged vison correlator as a function of the path length $|\overline{\Gamma}|$. 
  Each blue or red point represents the path-averaged value of the vison correlator, calculated over $500$
  randomly chosen paths with the same path length $|\overline{\Gamma}|$.  The hatched region shows the spread, which is the standard deviation of the correlator over 500 random paths of the same path length.
  }
  \label{fig:dimer-vison}
\end{figure}

\textit{(i) Dimer-dimer connected correlator.---}This correlator is defined as $C_{\mu\nu}^S = \langle \hat{\sigma}^z_{\mu} \hat{\sigma}^z_{\nu} \rangle - \langle \hat{\sigma}^z_{\mu} \rangle \langle  \hat{\sigma}^z_{\nu} \rangle$, where $\mu$ and $\nu$ are links of the Penrose tiling. 
Note that here we label the gauge fields $\hat{\sigma}$ by links $\mu$ and $\nu$, whereas we labeled the same operators by pairs of neighboring sites earlier.
We use Latin letters such as $i,j$ for sites, and Greek letters such as $\mu,\nu$ for links of the tiling.
We calculate this correlator for all pairs of links in $R_k$ and bin them according to the Euclidean distance between the links, where $k \in \{1, 2\}$. 
For each bin, we then calculate the average of the correlator, the Monte Carlo error in calculating the average of the correlator, and the standard deviation of the correlator in that bin. These are shown in~\cref{fig:dimer-vison}(a). 
We find that bin-averaged $C_{\mu \nu}^S$ decays exponentially with the distance between $\mu$ and $\nu$.
If our model were at a critical point, like the square or hexagonal lattice quantum dimer models at the RK point~\cite{fisher1961stastical,moessner2001shortranged,moessner2011introduction,moessner2001phase,schlittler2017phase}, then $C_{\mu \nu}^S$ would decay as a power-law.

\textit{(ii) Vison-vison correlator.---}We calculate the vison-vison correlator,  $C_{\overline{\Gamma}}^V = | \langle \hat{V}(\overline{\Gamma}) \rangle |$, for $10\,000$ randomly chosen paths in $R_k$. (Details of the path-selection algorithm are provided in the SM~\cite{supp}). 
The correlator is then averaged over those randomly chosen paths that have the same path length $|\overline{\Gamma}|$.
This path-averaged correlator is plotted in \cref{fig:dimer-vison}(b) as a function of the path length $|\overline{\Gamma}|$.
Now, $\hat{V}(\overline{\Gamma})$ depends on the entire path $\overline{\Gamma}$ and not just the path length.
Different paths with the same length result in a spread (nonzero standard deviation) in the correlator, represented by the hatched region in the figure.
We find that the path-averaged vison-vison correlator decays exponentially with the path length.
The correlation lengths are $5.73 \pm 0.08$ and $5.52 \pm 0.06$ in regions $R_1$ and $R_2$, respectively. (If the exponential decay is $\alpha e^{-x/\xi}$, then we refer to $\xi$ as the correlation length.)
[We also plot the path-averaged correlator as a function of the Euclidean distance between the endpoints of the paths in the SM (\cref{supp-fig:vison-ed}) and observe an exponential decay with distance.]
The top boundary of the hatched region shows that the exponential decay holds for the majority of individual paths, not just for the path-averaged correlator.
If the vison correlator were to approach a nonzero constant, we could conclude that the system was in an ordered phase.
In the absence of monomers, the exponential decay of the vison-vison correlator would indicate that the visons are not condensed, implying that the excitations of the gauge theory are fractionalized~\cite{senthil2000z2gauge}, as concluded in Refs.~\cite{balents2002fractionalization,ioselevich2002groundstate} for triangular lattice dimer models.
However, in the presence of  matter (monomers), this implication is not valid.
Open Wilson lines and closed Wilson loops become useful in such cases to determine the phase.


\textit{(iii) Open Wilson lines and closed Wilson loops.---}In gauge theories without matter, closed Wilson loops follow the area (perimeter) law in the confined (deconfined) phase~\cite{wilson1974confinement}.
However, in the presence of matter, Wilson loops follow the perimeter law in both phases. 
Despite this, the origin of the perimeter law differs between the two phases.
The distinct origins can be probed by studying the decay of open Wilson lines and closed Wilson loops.
In the confined phase, the perimeter law arises from fluctuations of the matter field, leading to the same decay rates for open Wilson lines and closed Wilson loops.
In contrast, in the deconfined phase, the perimeter law also receives contributions from fluctuations of the gauge field, causing closed Wilson loops to decay more slowly than open Wilson lines~\cite{fredenhagen1986,gregor2011diagnosing}.
See \cref{sec:gregor} of SM~\cite{supp} for details on the scaling of open Wilson lines and closed Wilson loops in the confined and deconfined phases.
Fredenhagen and Marcu~\cite{fredenhagen1983charged, fredenhagen1986confinement,fredenhagen1988dual} and later Gregor, Huse, Moessner, and Sondhi~\cite{gregor2011diagnosing} proposed using the ratio of open Wilson lines to closed Wilson loops to probe the origin of the perimeter law.
The numerical uncertainty in calculating this ratio can be large since both the numerator and the denominator decay exponentially with loop/path lengths. 
In this Letter, we instead study the scaling of open Wilson lines and closed Wilson loops without calculating their ratio.
Given two open paths $\Gamma_1$ and $\Gamma_2$ on the Penrose graph with the same endpoints, we define the following two quantities:
\begin{equation}
\label{eq:monomer-wislon}
    F_{\Gamma_1\Gamma_2}^{\text{op}} = \langle \hat{M} (\Gamma_1)  \rangle ~ \langle  \hat{M} (\Gamma_2)  \rangle~,\;\;\; F_{\Gamma}^{\text{cl}} = \langle \hat{M} (\Gamma)  \rangle, 
\end{equation}
where $\Gamma$ is a closed loop formed by connecting $\Gamma_1$ and $\Gamma_2$. 
$F_{\Gamma_1\Gamma_2}^{\text{op}}$ is a product of two open Wilson lines, while $F_{\Gamma}^{\text{cl}}$ is a closed Wilson loop, both corresponding to monomers.
Analogous to the vison correlator, $\langle \hat{M}(\Gamma) \rangle$ is referred to as the monomer correlator, where $\hat{M}(\Gamma) = \mathcal{P} \prod_{(k,l) \in \Gamma} \hat{\sigma}_{kl}^x$ for a closed loop $\Gamma$.
Similarly, given two paths, $\overline{\Gamma}_1$ and $\overline{\Gamma}_2$, on the dual graph with the same endpoints, we define the following two quantities:
\begin{equation}
    \label{eq:vison-wilson}G_{\overline{\Gamma}_1\overline{\Gamma}_2}^{\text{op}} = \lvert \langle \hat{V} (\overline{\Gamma}_1)  \rangle~  \langle \hat{V}(\overline{\Gamma}_2)  \rangle \rvert~,\;\;\; G_{\overline{\Gamma}}^{\text{cl}} = \lvert\langle \hat{V} (\overline{\Gamma}) \rangle \rvert~,
\end{equation}
where $\overline{\Gamma}$ is a closed loop formed by connecting $\overline{\Gamma}_1$ and $\overline{\Gamma}_2$.
$G_{\overline{\Gamma}_1\overline{\Gamma}_2}^{\text{op}}$ is a product of two open Wilson lines, while $G_{\overline{\Gamma}}^{\text{cl}}$ is a closed Wilson loop, both corresponding to visons.
We note that the Fredenhagen-Marcu order parameters for monomers and visons are $\sqrt{F_{\Gamma_1\Gamma_2}^{\text{op}} / F_{\Gamma}^{\text{cl}}}$ and $\sqrt{ G_{\overline{\Gamma}_1 \overline{\Gamma}_2}^{\text{op}} / G_{\overline{\Gamma}}^{\text{cl}}  }$ respectively.
In the limit of large loops, the expectation values of open Wilson lines and closed Wilson loops can be interpreted as follows: if the decay rates (with loop or path length) of closed Wilson loops and open Wilson lines corresponding to visons are the same, the visons are condensed, indicating an ordered phase.
Similarly, if closed Wilson loops and open Wilson lines corresponding to monomers  decay at the same rate, the monomers are condensed, signaling that the gauge theory is in the Higgs phase.

\begin{figure}[t]
  \centering
  \includegraphics[width=0.99\columnwidth]{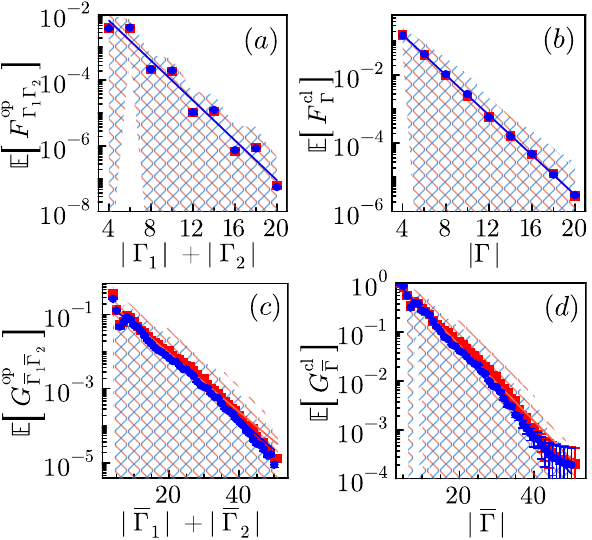}
  \caption{(a) and (b) show path-averages of  $F_{\Gamma_1\Gamma_2}^{\text{op}}$ and $F_{\Gamma}^{\text{cl}}$, while (c) and (d) show the path-averages of $G_{\overline{\Gamma}_1\overline{\Gamma}_2}^{\text{op}}$ and $G_{\overline{\Gamma}}^{\text{cl}}$. 
  Here, $\mathbb{E}$ denotes path-averaging over loops or paths on the Penrose graph or its dual. In (a), $\Gamma_1$ and $\Gamma_2$ are paths on the Penrose graph between the same endpoints while in (c), $\overline{\Gamma}_1$ and $\overline{\Gamma}_2$ are paths on the dual Penrose between the same endpoints.
  The legend for all four subplots is the same as the legend for \cref{fig:dimer-vison}.
  The red and blue points correspond to the averages in regions $R_1$ and $R_2$,  respectively.}
  
  \label{fig:fm-all}
\end{figure}

We now describe our calculation of open Wilson lines and closed Wilson loops corresponding to monomers. 
Approximately $18\,000$ and $9\,000$ closed loops are generated in regions $R_1$ and $R_2$, respectively, using a random walk (details of the algorithm are provided in the SM~\cite{supp}).
Each loop $\Gamma$ is then divided into two equal halves, $\Gamma_1$ and $\Gamma_2$, yielding a set of closed loops, $S_{M}^{\text{cl}}$, and a set of open path pairs, $S_{M}^{\text{op}}$.
Using Monte Carlo, we compute $F_{\Gamma_1 \Gamma_2}^{\text{op}}$ for all open path pairs $(\Gamma_{1},\Gamma_2)  \in S_{M}^{\text{op}}$ and $F_{\Gamma}^{\text{cl}}$ for all closed loops $\Gamma \in S_{M}^{\text{cl}}$.
Finally, $F_{\Gamma_1 \Gamma_2}^{\text{op}}$ is averaged over all pairs of open path pairs in $S_{M}^{\text{op}}$ with a given total path length $|\Gamma_1| + |\Gamma_2|$.
The path-averaged $F_{\Gamma_1 \Gamma_2}^{\text{op}}$ is plotted as a function of path length in~\cref{fig:fm-all}(a).
Following a similar path-averaging procedure for closed loops in $S_M^{\text{cl}}$, we obtain the path-averaged $F_{\Gamma}^{\text{cl}}$, as shown in~\cref{fig:fm-all}(b).
We find that both path-averaged $F_{\Gamma_1 \Gamma_2}^{\text{op}}$ and $F_{\Gamma}^{\text{cl}}$ 
decay exponentially with the path/loop length.
In $R_1$, the (unitless) correlation lengths are $1.43\pm 0.10$ for $F_{\Gamma_1 \Gamma_2}^{\text{op}}$ and $1.48 \pm 0.01$ for $F_{\Gamma}^{\text{cl}}$, while in $R_2$, they are $1.43\pm 0.11$ and $1.48 \pm 0.01$, respectively.
These values are equal within numerical accuracy, implying that path-averaged $F_{\Gamma_1 \Gamma_2}^{\text{op}}$ decays at the same rate as path-averaged $F_{\Gamma}^{\text{cl}}$.
Applying a similar path-averaging procedure to the vison Wilson lines and Wilson loops, we obtain the path-averaged $G_{\overline{\Gamma}_1\overline{\Gamma}_2}^{\text{op}}$ and $G_{\overline{\Gamma}}^{\text{cl}}$ plotted in~\cref{fig:fm-all}(c) and \cref{fig:fm-all}(d), respectively.
Both show an exponential decay with loop/path length in both regions, $R_1$ and $R_2$.
In $R_1$, the (unitless) correlation lengths of the exponential decay of $G_{\overline{\Gamma}_1\overline{\Gamma}_2}^{\text{op}}$ and $G_{\overline{\Gamma}}^{\text{cl}}$ are $5.66 \pm 0.26$ and $5.86 \pm 0.42$, respectively.
In $R_2$, they are $5.29 \pm 0.15$ and $5.30 \pm 0.19$, respectively. 
In both regions, the correlation lengths of path-averaged open Wilson lines and closed Wilson loops are equal within error.
Thus both $F_{\Gamma_1\Gamma_2}^{\text{op}}$ and $F_{\Gamma}^{\text{cl}}$  decay at the same rate, as do $G_{\overline{\Gamma}_1\overline{\Gamma}_2}^{\text{op}}$ and $G_{\overline{\Gamma}}^{\text{cl}}$. 
This implies that the gauge theory is in the confined phase, with visons and monomers condensed.


\textit{Discussion.---}We proposed  a frustration-free quantum monomer-dimer model on Penrose tilings, determined its exact ground state, mapped  the system to a $\mathbb{Z}_2$ gauge theory, and characterized the phase of the gauge theory at the frustration-free point.
Our analysis showed that both monomers and visons are condensed, indicating that the gauge theory is in the confined phase and the system is likely ordered.
We note that while the scaling behavior of Wilson loops and lines reveals the phase of the gauge theory, it does not reveal the specific type of ordering present in the system.
Whether the results of this Letter are specific to the particular Penrose tiling we studied, or whether they apply more generally to other Penrose tilings, is an important question which warrants a separate investigation.   
We leave it for future work.
While our results answer key questions about our model, they also open several new research directions. 
For instance, what type of ordering, in terms of dimers, does the system exhibit?
What is the phase of the system away from the RK point? 
By introducing additional terms in the Hamiltonian, monomers can be pinned to specific sites, and the density of mobile monomers can be varied (see \cref{sec:miscroscopic-spin-hamiltonian} of the SM~\cite{supp}).
Does the system undergo a phase transition as the mobile monomer density changes~\cite{syljuasen2005continuous}?
Does the ground state enter a deconfined phase upon introducing an additional density of monomers, analogous to what is seen in the square lattice~\cite{syljuasen2005continuous}?
Can a deconfined phase be obtained in a different quasicrystal with monomer membranes?
2D quasicrystals can be viewed as projections of higher-dimensional periodic lattices on certain planes~\cite{debruijn1981algebraic,debruijn1981algebraic2}. 
Which phenomena observed in quantum dimer models or gauge theories on higher-dimensional periodic lattices also occur in 2D quasicrystals?
Finally, how can neutral atoms, which can be arranged in arbitrary two-dimensional patterns using tweezer arrays~\cite{weimer2010rydberga,ebadi2021quantum,labuhn2016tunable,semeghini2021probing,barredo2016anatombyatom,samajdar2021quantum,zeng2025quantum}, and known to simulate dimer models on periodic lattices~\cite{semeghini2021probing,verresen2021prediction,zeng2025quantum}, be used to simulate quantum dimer models on quasicrystals?

\begin{acknowledgments}
\textit{Acknowledgments.---}This research was sponsored by the Army Research Office under Grant Number W911NF-23-1-0241, the National Science Foundation under Grant No. DMR-203715, and the NSF QLCI grant OMA-2120757. J.S.~and A.V.G.~were supported in part by NSF QLCI (award No.~OMA-2120757), AFOSR MURI, DoE ASCR Quantum Testbed Pathfinder program (awards No.~DE-SC0019040 and No.~DE-SC0024220), NSF STAQ program, ARL (W911NF-24-2-0107), DARPA SAVaNT ADVENT, and NQVL:QSTD:Pilot:FTL. J.S.~and A.V.G.~also acknowledge support from the U.S.~Department of Energy, Office of Science, National Quantum Information Science Research Centers, Quantum Systems Accelerator. J.S.~and A.V.G.~also acknowledge support from the U.S.~Department of Energy, Office of Science, Accelerated Research in Quantum Computing, Fundamental Algorithmic Research toward Quantum Utility (FAR-Qu).
\end{acknowledgments}

\bibliography{references}

\begin{thebibliography}{75}
\bibitem [{gau()}]{gauss-footnote}%
  \BibitemOpen
  \href@noop {} {}\bibinfo {note} {{N}ote that not all configurations of the
  $\hat{\sigma}$ and $\hat{\tau}$ fields satisfying the Gauss's law are valid
  maximal dimer coverings.}\BibitemShut {Stop}%
\bibitem [{vis()}]{vison-absolute-footnote}%
  \BibitemOpen
  \href@noop {} {}\bibinfo {note} {{T}he absolute value is needed because
  {$\langle \hat{V} (\overline{\Gamma}) \rangle$} may differ in sign for two
  paths that have the same endpoints for fixed monomers.}\BibitemShut {Stop}%
\end{thebibliography}
\clearpage

\widetext

\setcounter{equation}{0}
\setcounter{figure}{0}
\setcounter{table}{0}
\setcounter{page}{1}

\renewcommand{\theequation}{S\arabic{equation}}
\renewcommand{\thefigure}{S\arabic{figure}}

\normalsize\

\begin{center}
\textbf{\large A quantum monomer-dimer model on Penrose tilings}\vspace{0.2em}

\textbf{\large Supplemental Material}

\vspace{1em}

{\normalsize Jeet Shah\,\orcidlink{0000-0001-5873-8129},$^{1,2}$ Gautam Nambiar\,\orcidlink{0000-0003-4305-8600},$^{1}$ Alexey~V.~Gorshkov\,\orcidlink{0000-0003-0509-3421},$^{1,2}$ and Victor Galitski$^{1}$}\vspace{0.2em}

$^{1}$\textit{\small Joint Quantum Institute, NIST/University of Maryland, College Park, Maryland 20742, USA}\\
$^{2}$\textit{\small Joint Center for Quantum Information and Computer Science, NIST/University of Maryland, College Park, Maryland 20742, USA}

\vspace{1em}
\end{center}
\renewcommand{\thefigure}{S\arabic{figure}}
\setcounter{figure}{0}
\renewcommand\thesection{\arabic{section}}

In this supplemental material, we provide additional details on our calculation methods.
In \cref{sec:regions}, we describe the generation of the finite section of the Penrose tiling used in our Monte Carlo calculations.
In \cref{sec:monte-carlo}, we detail the Monte Carlo moves and the frequency of the number of times they are executed.
In \cref{sec:dimer-dimer-averaging}, we explain the bin-averaging procedure employed to produce the plot in \cref{fig:dimer-vison}(a) of the main text.
\Cref{sec:vison-avergaging} outlines our algorithm for generating random vison paths and the path-averaging procedure used for the plot in \cref{fig:dimer-vison}(b).
Next, in \cref{sec:fm-order-params-averaging}, we describe our algorithm for generating loops and paths for Wilson lines and loops, along with the associated path-averaging procedure.
We also comment on the behavior of open Wilson lines and closed Wilson loops in the deconfined and confined phases.
Finally, in \cref{sec:miscroscopic-spin-hamiltonian}, we provide a microscopic spin Hamiltonian that maps to a quantum monomer-dimer model on Penrose tilings in the low-energy limit. 
We also discuss additional terms that can be added to the Hamiltonian to vary the density of mobile monomers.

\section{Penrose tilings}\label{sec:regions}

In this section, we explain how we generate the finite piece of Penrose tiling on which we perform our Monte Carlo calculations.
Penrose tilings are fascinating, with many patterns hidden in them.
There are infinitely many Penrose tilings, and various methods are known for generating them, including matching rules, the inflation procedure, de Bruijn's cut-and-project method, and de Bruijn's pentagrid   method~\cite{debruijn1981algebraic,debruijn1981algebraic2}.
In this work, we use the inflation procedure to generate a finite section of a Penrose tiling for our Monte Carlo simulations.
\begin{figure}[htbp]
  \centering
  \includegraphics[width=0.6\textwidth]{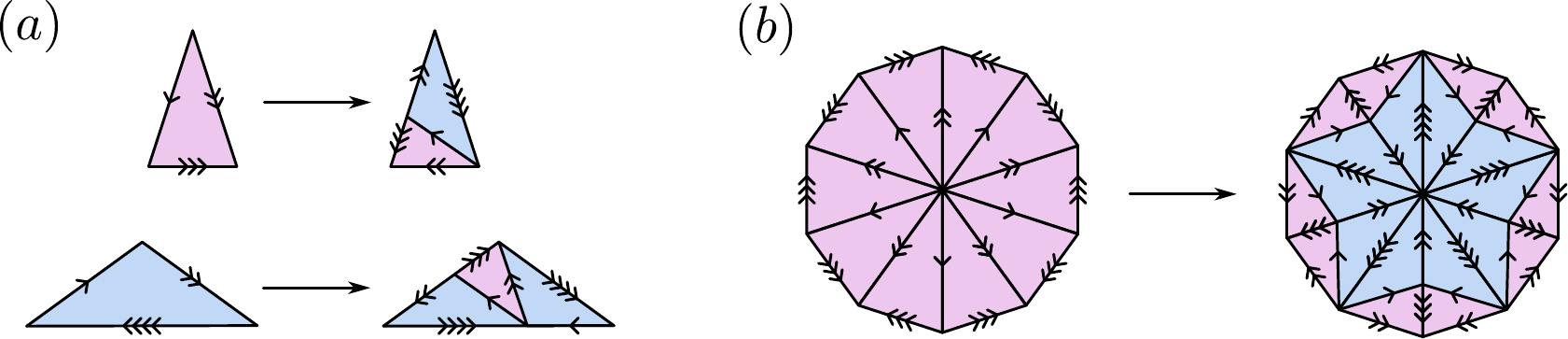}
  \caption{(a) Inflation rules. The two triangles on the left form the constituent shapes of the rhombuses in Penrose tilings. After inflation, they are divided into smaller triangles as shown. 
  (b) Starting with the configuration on the left, containing ten small triangles, we apply the inflation rules once to obtain the configuration on the right. Repeating the inflation process seven more times produces the  finite section of the tiling on which we perform our Monte Carlo calculations. This finite section of the tiling is shown in~\cref{supp-fig:regions}.}
\label{supp-fig:inflation}
\end{figure}
\begin{figure*}[htbp]
  \centering
  \includegraphics[width=0.9\textwidth]{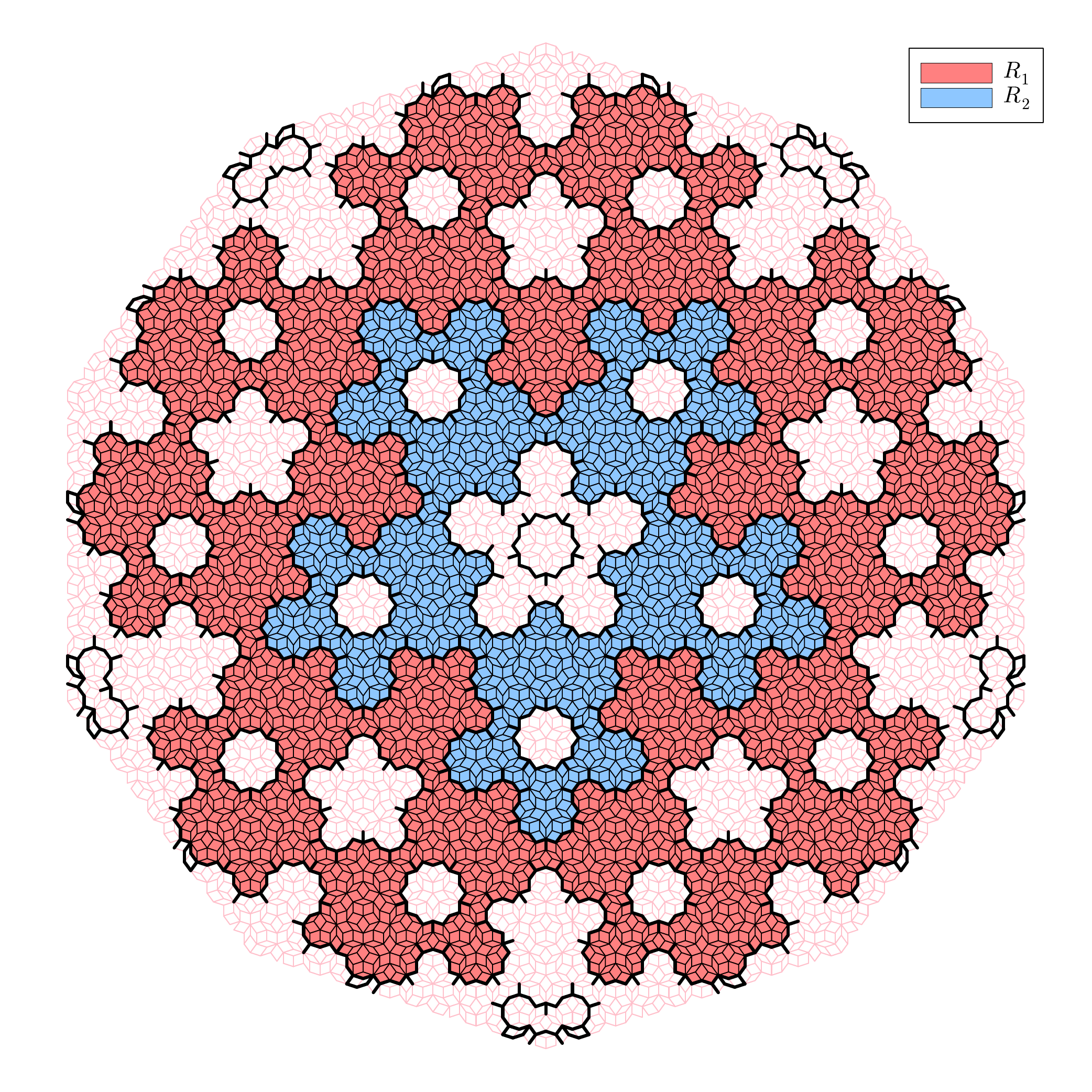}
  \caption{The finite section of the Penrose tiling used in our calculations. The thick black lines represent the monomer membranes, which divide the tiling into disconnected regions. $R_1$ (red) and $R_2$ (blue) denote the two regions where we compute various correlators.}
  \label{supp-fig:regions}
\end{figure*}
In the inflation procedure~\cite{grunbaum1987tilings}, a finite arrangement of triangles, as shown in~\cref{supp-fig:inflation}(a), is subdivided into smaller triangles similar to the original ones according to specific inflation rules. 
Two copies of the pink triangle of \cref{supp-fig:inflation}(a) form the thin rhombus of Penrose tilings. 
Similarly, two copies of the blue triangle in \cref{supp-fig:inflation}(a) form the thick rhombus of Penrose tilings.
Starting from the configuration on the left in \cref{supp-fig:inflation}(b), we recursively apply the inflation procedure eight times and remove all the links having three or four arrows on them, generating a tiling with $8\,111$ sites, as shown in~\cref{supp-fig:regions}. 
Since the process begins with a configuration that is five-fold rotationally symmetric, the resulting Penrose tiling in~\cref{supp-fig:regions} retains this five-fold rotational symmetry about its center.

\section{Monte Carlo details}\label{sec:monte-carlo}

In this section, we describe the Monte Carlo algorithm used to calculate correlators in the RK wavefunction.
We also provide details on the number of Monte Carlo moves performed and the number of samples collected.
To calculate the correlators in the RK wavefunction for the two regions, $R_1$ and $R_2$, shown in \cref{supp-fig:regions}, we employ the Metropolis Monte Carlo method at infinite temperature to sample maximal dimer coverings.
The Monte Carlo move we use for going from one maximal dimer covering to another is a monomer hop.
We  point out that the first term in~\cref{eq:dimer-monomer-H} implements a monomer hop.

Since the RK wavefunction factorizes into a tensor product of the RK wavefunctions on the regions $R_k$, it suffices to calculate correlators within one region at a time. 
In our calculation,  we first identify the monomer membranes, then remove the links belonging to the membranes
to obtain the graphs corresponding to regions $R_1$ and $R_2$. 
Region $R_1$ contains $3\,645$ sites, while region $R_2$ contains $1\,300$ sites.

\begin{figure}[htbp]
  \centering
  \includegraphics[width=0.45\columnwidth]{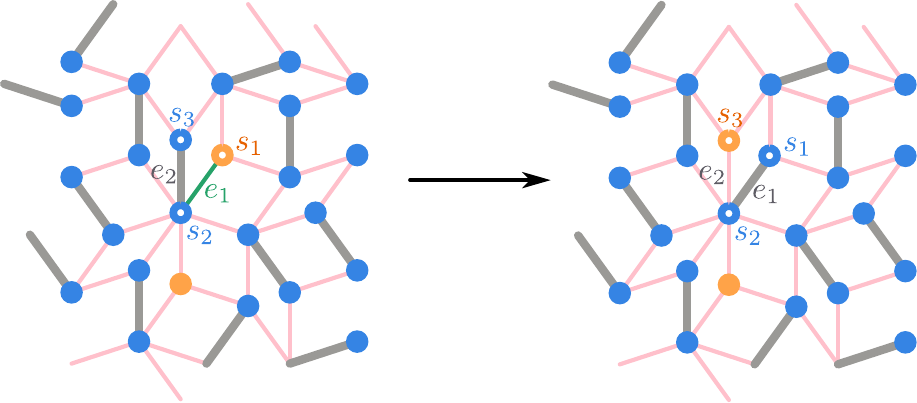}
  \caption{Hopping of a monomer from site $s_1$ to site $s_3$. Orange  circles and disks show monomers, while  blue circles with gray links between them represent dimers.}
  \label{fig:monomer-hop}
\end{figure}

At the beginning of the simulation, we find a maximal dimer covering of the chosen region $R_k$ using the Hopcroft-Karp algorithm~\cite{hopcroft1973an}.
Next, we perform a monomer hop, which we describe with the help of \cref{fig:monomer-hop}.
To execute a monomer hop, we first randomly choose a site (say $s_1$). If this site is not occupied by a monomer, then we do nothing. But if it is occupied by a monomer, 
 then we randomly select one of the edges (denoted as $e_1$) connected to $s_1$ and label the site at the other end of $e_1$ as $s_2$.
Since there's a monomer on $s_1$, there is no dimer on $e_1$ by the definition of a monomer.
Furthermore, $s_2$ cannot be occupied by a monomer since, in a maximal dimer covering, two neighboring sites cannot both be occupied by monomers.
Next, we identify the edge extending from $s_2$ that has a dimer on it (labeled  by $e_2$) and label the site at the other end of $e_2$ as $s_3$.
The monomer then hops from $s_1$ to $s_3$, and the dimer on $e_2$ shifts to $e_1$, completing the monomer move.

Each Monte Carlo step consists of  a monomer move. 
We note that monomer moves  are ergodic~\cite{flicker2020classical} and sufficient for the convergence of the Monte Carlo calculation.
We sample the correlators every $2\,000$ Monte Carlo steps for $R_1$ and every $5\,000$ steps for $R_2$ to reduce correlations between samples.
For each region we average the correlators over at least $2\times 10^6$ maximal dimer coverings. 

\section{Averaging of the dimer-dimer correlator}
\label{sec:dimer-dimer-averaging}

In this section, we describe the post-processing and the averaging procedure used to calculate the bin-averaged dimer-dimer correlator plotted  in~\cref{fig:dimer-vison}(a) of the main text.

In a Monte Carlo run, we calculate $\langle \psi_{\text{RK}} | \hat{\sigma}_{\mu}^z | \psi_{\text{RK}} \rangle$ for all links $\mu$, and  $\langle \psi_{\text{RK}} | \hat{\sigma}_{\mu}^z\hat{\sigma}_{\nu}^z | \psi_{\text{RK}} \rangle$ for all pairs of links $(\mu,\nu)$ by averaging over the $2\times 10^6$ dimer coverings generated in the Monte Carlo run. 
We perform 20 such Monte Carlo runs.
Using this information, we then calculate $C_{\mu\nu}^S = \langle \hat{\sigma}^z_{\mu} \hat{\sigma}^z_{\nu} \rangle - \langle \hat{\sigma}^z_{\mu} \rangle \langle  \hat{\sigma}^z_{\nu} \rangle$ for all pairs of links $(\mu,\nu)$.
We also calculate the Euclidean distances $d_{\mu \nu}$ between the midpoints of all pairs of links.
If we plot $C_{\mu \nu}^S$ vs.~$d_{\mu \nu}$, we do not get a smooth curve; rather we get a scatter of points because of the irregularity of the Penrose tilings.
To interpret the dimer-dimer connected correlator as a function of the distance, we bin the links according to $d_{\mu \nu}$, where the bins have a width of $0.4a$ ($a$ is the side length of the rhombuses of the tiling).
More explicitly, the first bin is a set of all pairs of links $(\mu, \nu)$ where the distance between the links $\mu$ and $\nu$ is between $0$ and $0.4a$. 
The second bin is a set of all pairs of links where the distance between the links is between $0.4a$ and $0.8a$, and so on.
Then we define the bin-averaged dimer-dimer correlator for bin $b$ as 
\begin{equation}
    \overline{A}_b = \frac{1}{N_b} \sum_{(\mu, \nu) \in b} \left[ \langle \psi_{\text{RK}} | \hat{\sigma}_{\mu}^z \hat{\sigma}_{\nu}^z | \psi_{\text{RK}} \rangle  - \langle \psi_{\text{RK}} | \hat{\sigma}_{\mu}^z | \psi_{\text{RK}} \rangle  \langle \psi_{\text{RK}} |  \hat{\sigma}_{\mathsf{\nu}}^z | \psi_{\text{RK}} \rangle\right]~. 
\end{equation}
For each bin $b$, we calculate the average and the standard deviation of $C_{\mu\nu}^S$, denoted by $\overline{A}_{b}$ and $\Delta  A_{b}$, respectively.
The bin averages, $\overline{A}_b$, are plotted as the data points in \cref{fig:dimer-vison}(a).
The top and bottom boundaries of the hatched regions in \cref{fig:dimer-vison}(a) of the main text are $\overline{A}_b + \Delta A_b$ and $\overline{A} - \Delta A_b$, respectively.
They look asymmetrical about the data points in the plot because of the logarithmic scale used there.
Thus, the hatched regions indicate the spread of the correlator about its average value.
The error bars in the plot represent the standard error in the average of the bins, $\overline{A}_b$, arising from a finite number of Monte Carlo samples.
The standard error is determined from the 20 independently done Monte Carlo runs.
The standard error is expected to be proportional to $1/\sqrt{\text{Number of samples}}$. 
For the dimer-dimer connected correlator, the number of samples we take is $20\times 2 \times 10^6$, which will lead to a standard error of the order of $1.6 \times 10^{-4}$.
Thus correlator values below this level in Fig.~\ref{fig:dimer-vison}(a) cannot be trusted.
The saturation of the connected correlator for distances beyond $\approx 7a$ occurs because of the finite number of Monte Carlo samples and not due to the underlying physics.
We note that we have verified that using a bin width different from $0.4a$, such as $0.1a, 0.2a$, or  $0.5a$, does not change our results. 
We still obtain an exponential decay of the bin-averaged correlator, with correlation lengths that are equal, within error margins, to those obtained for a bin width of $0.4a$.
\begin{figure}[htbp]
  \centering
  \includegraphics[width=0.45\columnwidth]{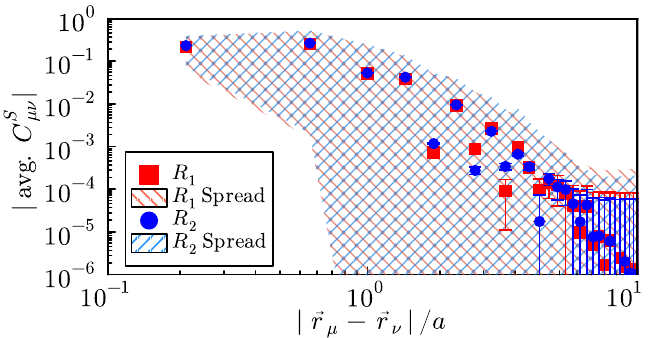}
  \caption{Log-log plot of the dimer-dimer connected correlator, previously shown on a log-scale in~\cref{fig:dimer-vison}(a). 
  The style of the data points, error bars, hatched regions, and legend are identical to those in~\cref{fig:dimer-vison}(a).
  The data points as well as the upper boundary of the hatched regions do not seem to follow a straight line, thereby ruling out power-law decay of the correlations.
  }
  
  \label{supp-fig:dimer-log-log}
\end{figure}
Moreover, to rule out a power-law decay indicative of criticality, we replot the dimer-dimer connected correlator on a log-log scale in \cref{supp-fig:dimer-log-log}.
Comparing the data points in this figure with log-scale plot of \cref{fig:dimer-vison}(a), we see that the data is far more consistent with an exponential than a power-law decay.
Also, the shape of the upper boundary of the hatched regions in both figures supports an exponential decay of the correlator rather than a power-law decay.

\begin{figure*}[htbp]
  \centering
  \includegraphics[width=0.8\textwidth]{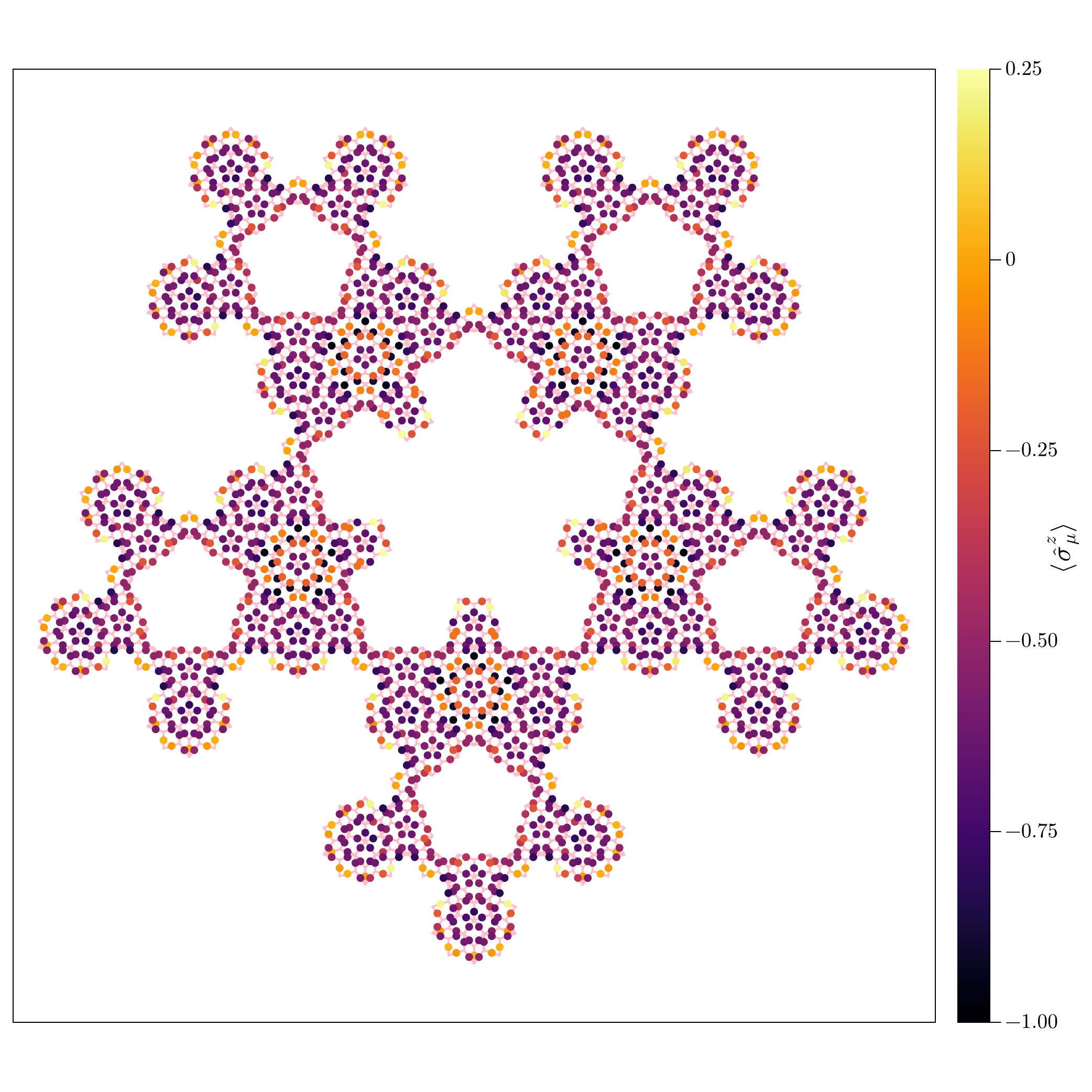}
  \caption{Color plot of $\hat{\sigma}^z_\mu$ in the region $R_2$ of the Penrose tiling. 
  The image reveals that the five-fold rotational symmetry is preserved in the RK wavefunction. 
  Spins are located on the edges $\mu$ of the graph.}
  \label{supp-fig:R3-sz-mean}
\end{figure*}

We also plot $\langle \hat{\sigma}^z_\mu \rangle$ for each spin $\mu$ in the $R_2$ region in \cref{supp-fig:R3-sz-mean}, and see that the plot has five-fold rotational symmetry, indicating that the rotational symmetry is not broken.

\section{Averaging of the vison correlator}
\label{sec:vison-avergaging}

\begin{figure*}[htbp]
  \centering
  \includegraphics[width=0.75\textwidth]{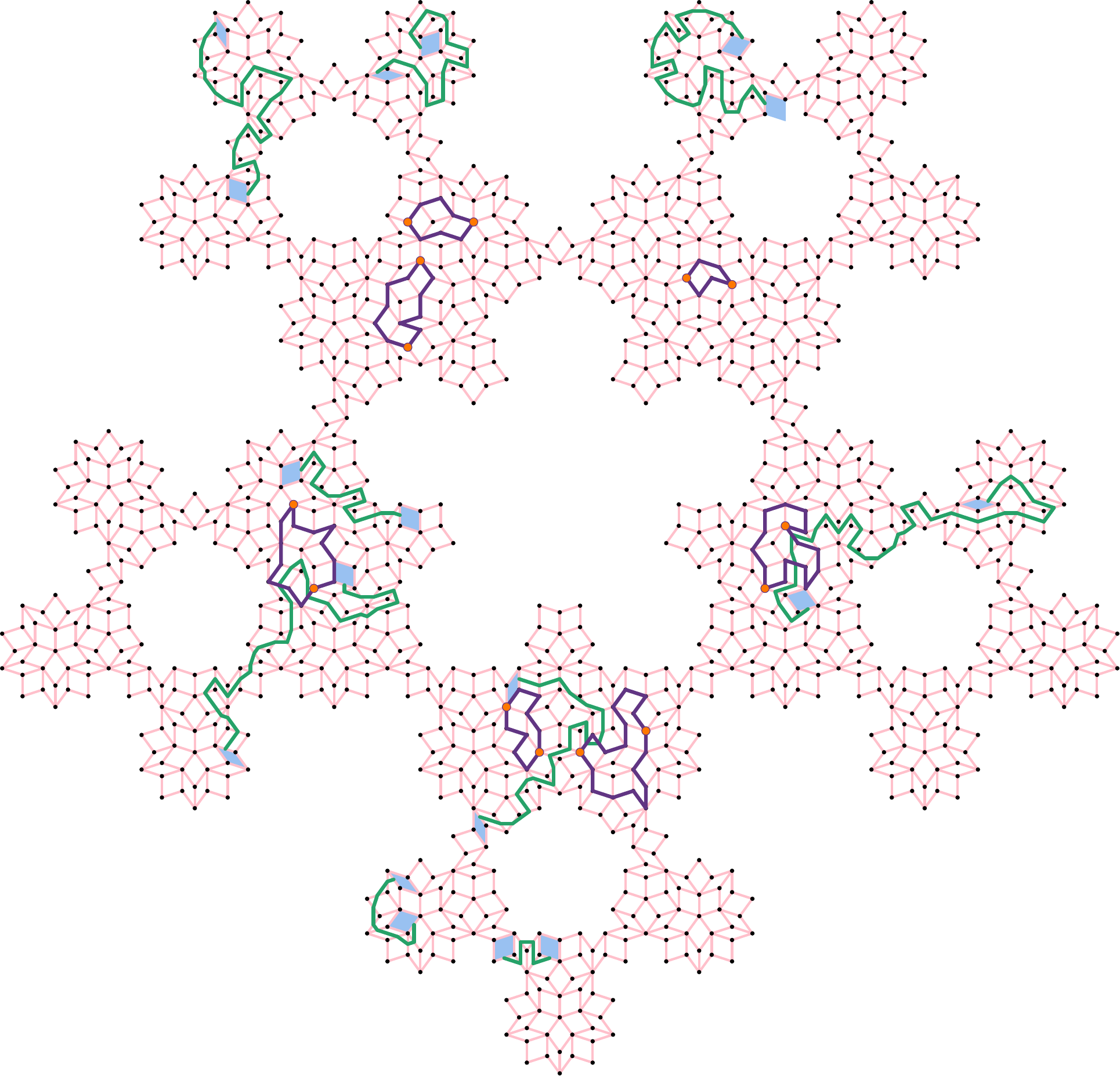}
  \caption{A few examples of vison paths and monomer FM loops in $R_2$ are shown, which  we use in our calculation. The green lines represent vison paths, while the blue plaquettes mark the endpoints where the visons are created. Each green path crosses every plaquette an even number of times, except for the blue plaquettes at the ends. The purple loops correspond to the loops for the monomer FM order parameter. The orange sites along these loops indicate points where the loops are split to create two open paths of equal length. The monomer operator along one such open path transports monomers between its orange-colored endpoints if it is permitted by the maximal dimer covering constraint.}
  \label{supp-fig:path-examples}
\end{figure*}

In this section, we provide details on the random selection of vison paths, the evaluation of the vison correlator, and path averaging used to generate \cref{fig:dimer-vison}(b) in the main text. 
We first describe our algorithm for generating a valid vison path. 
These details are included for completeness, although we do not believe that our results are specific to the vison paths chosen through our algorithm.
We verify that considering half as many paths still gives similar plots, indicating that the number of paths chosen is large enough for the path averages to have converged.
Recall that a vison path $\overline{\Gamma}$ is a path on the dual graph of the Penrose tiling. 
Equivalently, vison paths can be thought of as paths through the plaquettes of the Penrose tiling.
These paths cross an even number of edges of all the plaquettes except for  plaquettes $P$ and $Q$, corresponding to the endpoints of the path. 
Furthermore, the path crosses only one edge of $P$ and only one edge of $Q$.
\Cref{supp-fig:path-examples} shows some of the vison paths we consider.
\begin{figure}[htbp]
  \centering
  \includegraphics[width=0.45\columnwidth]{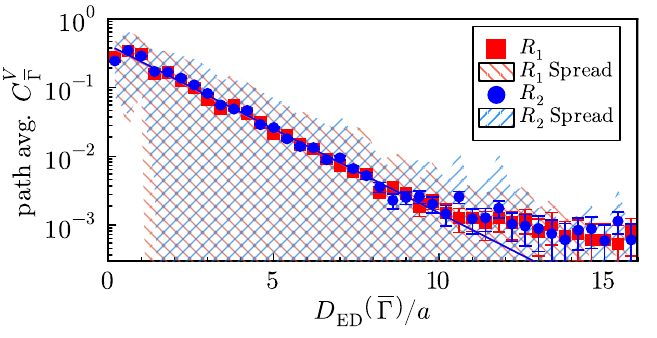}
  \caption{The  path-averaged vison correlator is plotted as a function of the Euclidean distance between the first and the last links of paths, denoted by $D_{\text{ED}}(\overline{\Gamma})$ divided by $a$.
  We group the paths $\overline{\Gamma}$ into bins of width $0.4a$ based on $D_{\text{ED}}(\overline{\Gamma})$.
  The data points represent  the average of the vison correlator over paths in each bin.
  The hatched regions indicate the standard deviations of the vison correlator within each bin, while the error bars represent the standard error in the Monte Carlo calculation of the bin averages.}
  \label{supp-fig:vison-ed}
\end{figure}
These paths are generated as follows.
We start from a randomly chosen link/edge, say $e_1$. 
This link will be shared between two plaquettes, say $P_1$ and $P_2$.
Next, we randomly choose a link, say $e_2$, from either $P_1$ or $P_2$ and add it to the path.
We repeat these steps, ensuring at each step that we do not backtrack and that no more than two links of a plaquette are visited at any intermediate step.
This guarantees that we generate valid vison paths.

We repeat the entire process until we get 500 paths for each of the path lengths $2, 4, 6, \ldots, 40$ in each of the regions $R_1$ and $R_2$.
Finally, we use the five-fold rotational symmetry and the reflection symmetry to generate ten symmetry-related paths for every path. 
All the symmetry-related paths have the same value of the vison correlator in principle.
However, because of the finite number of samples used in the Monte Carlo calculation, the symmetry-related paths give slightly different values of the correlator.
We average $\langle \hat{V} (\Gamma) \rangle$ (without taking the absolute value) over these ten symmetry-related paths to increase the accuracy of our calculation.
We sample the vison correlator for all these paths over $2\times 10^6$ maximal dimer coverings generated during the Monte Carlo calculation.
This gives us the vison correlator for all the randomly generated paths.

Now we explain the path-averaging procedure. 
Recall that the vison correlator $|\langle \hat{V} (\overline{\Gamma}) \rangle|$ is a function of not just the path length $|\overline{\Gamma}|$ (the number of links $\overline{\Gamma}$ intersects), but the entire path $\overline{\Gamma}$. 
To interpret the vison correlator as a function of the path length, we calculate the average of the vison correlator for all paths we generate that have the same path length.
This is the path-averaged vison correlator plotted in~\cref{fig:dimer-vison}(b).
For a given path length, we also calculate the standard deviation of the vison correlator over paths of that given length.
The hatched regions in \cref{fig:dimer-vison}(b) represent these standard deviations. 
We compute the standard error of the vison correlator for a particular path $\overline{\Gamma}$ from the standard deviation of the vison correlator of the ten paths that are symmetry-related to $\overline{\Gamma}$ (one of these ten paths is $\overline{\Gamma}$ itself). 
We average these standard errors over paths of the same length to obtain the standard error of the path-averaged vison correlator---that is, the average of the errors gives the error of the average. We calculate the standard error of path-averages in a similar way in~\cref{fig:dimer-vison}(b),~\cref{fig:fm-all}, and~\cref{supp-fig:vison-ed}.
The error bars in  these figures represent this standard error.

We also plot the path-averaged vison correlator as a function of the Euclidean distance between the first and last links along the paths in $R_1$ and $R_2$; see~\cref{supp-fig:vison-ed}.
We denote this Euclidean distance for a given path $\overline{\Gamma}$ by $D_{\text{ED}}(\overline{\Gamma})$.
Here, we employ a binning procedure similar to that used for the dimer-dimer correlator in~\cref{sec:dimer-dimer-averaging}. 
Specifically, we bin the paths based on  their $D_{\text{ED}}(\overline{\Gamma})$ values, using the same bin width as that chosen for the dimer-dimer correlator, which is $0.4a$.
For each bin $b$, we compute the average and standard deviation of the correlator over all paths within $b$.
Additionally, we calculate the standard error in the average correlator for each bin, arising from the finite precision of the Monte Carlo calculation.
We find that the path-averaged vison correlator decays exponentially with the Euclidean distance between the endpoints, with correlation lengths of $1.79a \pm 0.03a$ and $1.78a \pm 0.04a$ in $R_1$ and $R_2$, respectively.
Note that $D_{\text{ED}}(\overline{\Gamma})$  is bounded by a constant times the path length $| \overline{\Gamma} |$, which represents the number of links crossed by $\overline{\Gamma}$. 
Thus, the exponential decay of the vison correlator with the path length implies an exponential decay with the Euclidean distance between the first and last links of the paths. 
Since we sample over $2\times 10^6$ maximal dimer coverings to calculate the vison correlator for each path, the statistical accuracy of our Monte Carlo calculation is expected to be on the order of $1/\sqrt{2\times 10^6} \approx 7 \times 10^{-4}$.
Therefore, correlator values below this level are not reliable.
In Fig.~\ref{supp-fig:vison-ed}, the vison correlator appears to saturate at a level of approximately $10^{-3}$ at large distances.
This is on the order of the expected precision of our calculation, indicating that the saturation is an artifact due to the finite number of Monte Carlo samples, rather than a physical phenomenon.


\section{Averaging of the  Wilson loops}
\label{sec:fm-order-params-averaging}

In this section, we discuss the behavior of open Wilson lines and closed Wilson loops in the confined and deconfined phases. We also provide details of the path-averaging procedure for computing the path-averages of open Wilson lines and closed Wilson loops.

\subsection{Behavior of the Wilson lines and loops}
\label{sec:gregor}

In this section, we comment on the behavior of open Wilson lines and closed Wilson loops in the deconfined and the confined phases based on the results of Ref.~\cite{gregor2011diagnosing}.
Here we only describe the results, the details of which can be found in Sec.~III of Ref.~\cite{gregor2011diagnosing}.

In a pure gauge theory without matter fields, closed Wilson loops decay following an area law in the confined phase and a perimeter law in the deconfined phase, both arising from gauge field fluctuations.
Now, in the presence of heavy dynamical matter with a matter coupling $J$ and a gauge coupling $K$, closed Wilson loops are screened by matter fields.
As a result, the Wilson loop corresponding to a closed loop $\mathcal{L}$ decays as $\alpha_{J}^{|\mathcal{L}|}$ in the confined phase.
Here, $|\mathcal{L}|$ denotes the number of gauge variables along the loop $\mathcal{L}$ and $\alpha_J < 1$.
On the other hand, in the deconfined phase, closed Wilson loops decay as $\beta_{K}^{|\mathcal{L}|}$, where $\beta_{K} < 1$ depends on the gauge field strength $K$.
This distinction makes it clear that, while in the confined phase the perimeter law arises from matter field fluctuations, in the deconfined phase, it originates from gauge field fluctuations.
Now, consider open Wilson lines in the two phases. 
In the confined phase, matter fields are expected to screen open Wilson lines, leading to a decay of $\alpha_J^{|\mathcal{C}|}$ for an open path $\mathcal{C}$.
However, in the deconfined phase, open Wilson lines decay as $\alpha_{J}^{g} \beta_K^{g + |\mathcal{C}|}$, where $g$ is the distance between the endpoints of $\mathcal{C}$.
This indicates that, in the deconfined phase, both matter and gauge field fluctuations contribute to the exponential decay of open Wilson lines.
Ref.~\cite{gregor2011diagnosing} proposed using the ratio of open to closed Wilson loops as an elegant way to determine whether the decay is caused solely by matter field fluctuations or by both matter and gauge field fluctuations.
This ratio is known in the literature as the Fredenhagen-Marcu (FM) order parameter~\cite{fredenhagen1983charged,fredenhagen1986confinement,fredenhagen1988dual}.
Since open Wilson lines decay faster than closed Wilson loops in the deconfined phase, the FM order parameter tends to zero in the limit of large loops.
In contrast, in the confined phase, where open Wilson lines and closed Wilson loops decay at the same rate, the FM order parameters approaches a nonzero constant.
In our work, rather than studying the ratio of open Wilson lines to closed Wilson loops, we analyze their decay rates separately.

\subsection{Path averaging procedure}

We now  describe our algorithms for obtaining Wilson lines and loops corresponding to monomers and visons.

The monomer Wilson lines and loops  are defined on the Penrose graph, while the vison Wilson loops and lines  are defined on its dual.
To generate a loop on the Penrose graph within a given region $R_k$, we randomly select a site and perform a non-backtracking random walk from that point until the walk intersects itself. 
Since Penrose tiling forms a bipartite graph, all resulting loops have even lengths.
Next, we divide each loop $\Gamma$ at random locations into two equal-length paths, $\Gamma_1$ and $\Gamma_2$.
We also ensure that all loops are unique, meaning each loop is counted only once in the path averages.
\Cref{supp-fig:path-examples} illustrates  some of the loops in region $R_2$ that we consider.
We then exploit the five-fold rotational symmetry of the tiling, as well as reflection symmetry about a vertical axis through the center, to generate ten symmetry-related loops.
Similarly, we generate ten symmetry-related paths for each open path.
These additional copies improve numerical accuracy and allow estimation of error bars for the monomer correlator corresponding to open paths and closed loops.
At the end of this procedure, we obtain a set of closed loops $S_{M}^{\text{cl}}$ and a set $S_{M}^{\text{op}}$ consisting of pairs of open paths,   such that each pair forms a closed loop.

In our numerical analysis, we select between 300 and $5\,000$ loops in $R_1$ and between 100 and $2\,700$ loops in $R_2$. 
Following the approach used in \cref{sec:vison-avergaging}, for a given path length $l$, we compute the average and standard deviation of  $F_{\Gamma_1\Gamma_2}^{\text{op}}$, for all pairs of paths $(\Gamma_1, \Gamma_2)$ in $S_M^{\text{op}}$ satisfying  $ |\Gamma_1| + |\Gamma_2| = l$.
This path-averaged value of $F_{\Gamma_1\Gamma_2}^{\text{op}}$ is represented by the data points  in~\cref{fig:fm-all}(a) of the main text, while 
the red and blue hatched regions in the same figure indicate the standard deviation of  $F_{\Gamma_1\Gamma_2}^{\text{op}}$ for fixed $|\Gamma_1| + |\Gamma_2|$.
The error bars in~\cref{fig:fm-all}(a) represent the standard error in the path-averaged monomer correlator, arising from the finite number of steps in the Monte Carlo calculation.
This error in the path-averaged value of $F_{\Gamma_1\Gamma_2}^{\text{op}}$ is calculated by averaging the standard error in $F_{\Gamma_1\Gamma_2}^{\text{op}}$ which arises due to the finite number of Monte Carlo steps, over all pairs of paths $(\Gamma_1, \Gamma_2)$ with a fixed total path length $|\Gamma_1| + |\Gamma_2|$.
Similarly,~\cref{fig:fm-all}(b) displays the path-averaged value of $F_{\Gamma}^{\text{cl}}$, along with its standard error.
This standard error is calculated by averaging the standard error in $F_{\Gamma}^{\text{cl}}$ over loops of a given length.
It also shows the standard deviation of $F_{\Gamma}^{\text{cl}}$ over closed loops of fixed length by the red and blue hatched regions.
We note that Ref.~\cite{gregor2011diagnosing}, in its Sec.~8.3, suggests that the (monomer) Wilson loop is defined as a dimer rearrangement on a loop that respects the hard-core dimer constraint. 
Such a constraint allows for the possibility of creating additional monomers.
Our Wilson loop, corresponding to monomers, $\langle \hat{M} (\Gamma) \rangle$, differs from this in that we impose the maximal dimer covering constraint, which is more restrictive than the hard-core dimer constraint [see definition of $\hat{M} (\Gamma)$ in \cref{eq:monomer-op}].
However, since we always calculate expectation values in the RK wavefunction, we obtain the same result even if we impose the weaker hard-core dimer constraint instead of the maximal dimer covering constraint.
We also note that path-averaged $F_{\Gamma_1 \Gamma_2}^{\text{op}}$, shown in \cref{fig:fm-all}(a), exhibits an alternating feature whose origin remains unclear. The data points for loop lengths $4, 8, 12, \ldots$ appear to follow a different curve than those for loop lengths $6, 10, 14, \ldots$. 

Having described the details of calculating path-averaged Wilson lines and loops corresponding to monomers,  we now provide the corresponding details for calculating path-averaged Wilson lines and loops corresponding to visons.
Random loops for calculating the path-averages of $G_{\overline{\Gamma}_1 \overline{\Gamma}_2}^{\text{op}}$ and $G_{\overline{\Gamma}}^{\text{cl}}$ are generated using essentially the same procedure as that used to generate paths for the vison correlator $C_{\overline{\Gamma}}^V$ in \cref{sec:vison-avergaging}.
The only difference is that we continue the random walk until the path intersects itself.
Using this procedure, we generate loops $\overline{\Gamma}$ on the dual graph with lengths ranging from 4 to 51 for $R_1$ and from 4 to 50 for $R_2$.
For each of these lengths, the number of loops ranges from about 100 to $2\,200$  in the two regions. 
We denote the set of loops generated by $S_V^{\text{cl}}$.
Next, we split each closed loop $\overline{\Gamma}$ into two segments, $\overline{\Gamma}_1$ and $\overline{\Gamma}_2$, of length $|\overline{\Gamma}|/2$ or $(|\overline{\Gamma}| \pm 1) /2$, depending on whether $|\overline{\Gamma}|$ is odd or even.
This process yields a set of pairs of open paths $(\overline{\Gamma}_1,\overline{\Gamma}_2)$ on the dual graph, denoted by $S_{V}^{\text{op}}$.

For every loop and path, we also generate ten symmetry-related counterparts (accounting for five-fold rotational and reflection symmetries)  to improve numerical accuracy.
Next, we compute the expectation value of the vison operator $\langle \hat{V} (\overline{\Gamma}) \rangle$ for all the randomly generated paths and loops over the $2\times 10^6$ maximal dimer coverings produced during the Monte Carlo simulation.
To calculate path-averaged $G_{\overline{\Gamma}_1\overline{\Gamma}_2}^{\text{op}}$, we compute $\lvert \langle \hat{V} (\overline{\Gamma}_1) \rangle \langle \hat{V} (\overline{\Gamma}_2) \rangle \rvert$ for all pairs $(\overline{\Gamma}_1,\overline{\Gamma}_2) \in S_{V}^{\text{op}}$ and average the results over all pairs with a fixed total length $|\overline{\Gamma}_1| + |\overline{\Gamma}_2|$, denoted by $\overline{l}$.
This gives us  path-averaged $G_{\overline{\Gamma}_1\overline{\Gamma}_2}^{\text{op}}$  as a function of loop length $\overline{l}$ which is plotted in \cref{fig:fm-all}(c).
Similarly, to compute path-averaged $G_{\overline{\Gamma}}^{\text{cl}}$, we evaluate $\lvert \langle \hat{V} (\overline{\Gamma}) \rangle \rvert$ for all closed loops $\overline{\Gamma} \in S_{V}^{\text{cl}}$ and average the results over all loops with fixed length $|\overline{\Gamma}|$.
The resulting path-averaged $G_{\overline{\Gamma}}^{\text{cl}}$ as a function of loop length is plotted in \cref{fig:fm-all}(d).
The error bars and the spread in $G_{\overline{\Gamma}_1 \overline{\Gamma}_2}^{\text{op}}$ and $G_{\overline{\Gamma}}^{\text{cl}}$ are determined using a procedure similar to that used for $F_{\Gamma_1 \Gamma_2}^{\text{op}}$ and $F_{\Gamma}^{\text{cl}}$ described above.

Note that we do not compute the Fredenhagen-Marcu (FM) order parameters, given by  $\sqrt{F_{\Gamma_1 \Gamma_2}^{\text{op}} / F_{\Gamma}^{\text{cl}}}$ and $\sqrt{ G_{\overline{\Gamma}_1 \overline{\Gamma}_2}^{\text{op}} / G_{\overline{\Gamma}}^{\text{cl}}}$, because their numerators and denominators are exponentially small quantities.  
Even a small error arising from the finite number of Monte Carlo steps would be amplified in calculating the ratio.
Additionally, certain closed loops never exhibit an alternating configuration of dimer and no-dimer in any of the $2\times 10^6$ maximal dimer configurations generated during the Monte Carlo calculation.
The structure of the Penrose tiling could be responsible for preventing these loops from having an alternating dimer and no-dimer configuration.
This implies that the monomer correlator---and hence the denominator of the FM order parameter corresponding to monomers---for such loops is exactly zero.
For these loops, the monomer FM order parameter is not well-defined.
Comparing the exponential decay rates of open Wilson lines and closed Wilson loops separately circumvents this issue of the FM order parameter being undefined and directly probes the origin of the perimeter law in Wilson loops.

\section{Microscopic spin Hamiltonian}
\label{sec:miscroscopic-spin-hamiltonian}

\begin{figure}[htbp]
  \centering
  \includegraphics[width=0.5\textwidth]{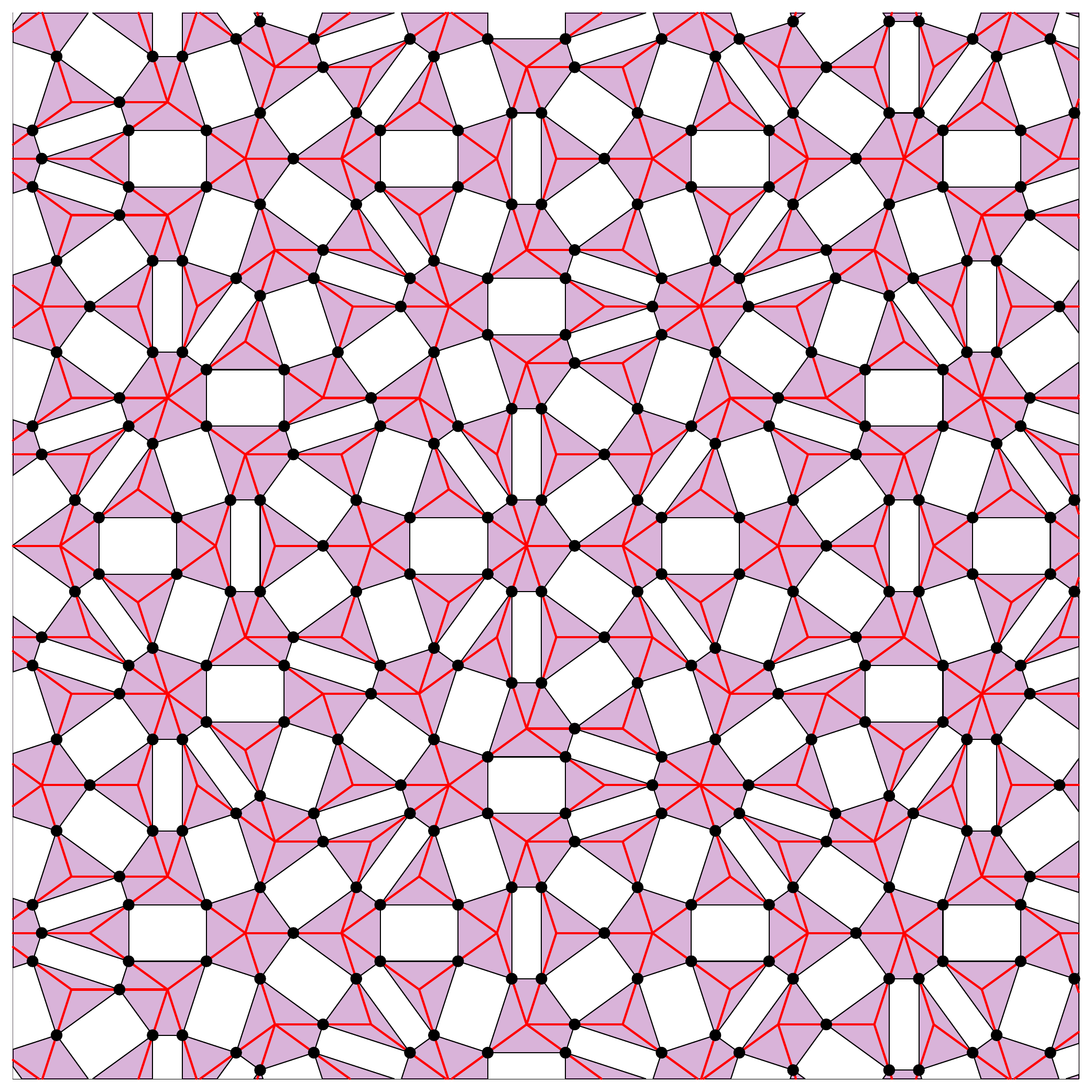}
  \caption{Line graph of the Penrose tiling used to define our microscopic spin Hamiltonian, $\hat{H}_{\text{spin}}$. 
  The red edges form the Penrose tiling. 
  Spin-1/2s are located at the midpoints of the red edges and are represented by black dots. 
  These dots are the vertices of the line graph of the Penrose tiling.
  The black edges are the edges of the line graph.
  The line graph can be viewed as a graph of corner-sharing polygons (shaded in purple).
  Note that there is a one-to-one correspondence between the shaded polygons and the sites of the Penrose tiling.}
  \label{supp-fig:penrose-poly}
\end{figure}

In this section, we provide a microscopic spin Hamiltonian that maps, in the low-energy limit, to a quantum monomer-dimer model on Penrose tilings. We do this using the techniques of Ref.~\cite{balasubramanian2022classical}.

The low-energy subspace consists of maximal dimer coverings. 
Although the effective Hamiltonian obtained here differs from the Hamiltonian in \cref{eq:dimer-monomer-H} studied in the main text, it offers a concrete method for realizing a quantum monomer-dimer model starting from spins.

We begin by constructing the line graph, made of corner-sharing polygons, as illustrated in \cref{supp-fig:penrose-poly}, derived from a Penrose tiling under consideration. 
This graph is defined as follows: each vertex of the line graph represents an edge of the Penrose tiling.
There is an edge between two vertices of the line graph if and only if the corresponding edges of the Penrose tiling share a site.
After constructing the line graph, we identify polygons whose vertices are midpoints of edges sharing a common site on the Penrose tiling.
These polygons are shaded in purple in \cref{supp-fig:penrose-poly}. 
For a polygon $p$, we label its number of vertices by $q_p$.

In the microscopic Hamiltonian, spin-1/2s $\hat{S}_{\mu}$ are located at the midpoints of the edges $\mu$ of the Penrose tiling (vertices of the line graph). 
These coincide with the gauge fields $\hat{\sigma}$ defined in the main text.
The microscopic spin Hamiltonian is given by
\begin{equation}\label{eq:microscopic-H}
\hat{H}_{\text{spin}} = \underbrace{V \sum_{\text{\color{purple}{polygons} }\textcolor{purple}{p}} \left( \hat{S}^z_p + \frac{q_p}{2} - h \right)^2}_{\hat{H}_0} + \underbrace{\Omega \sum_{\mu} \hat{S}^x_\mu}_{\hat{H}_1}, 
\end{equation}
where the first sum is over all purple-colored polygons $p$ in \cref{supp-fig:penrose-poly}, $\hat{S}_p^z = \sum_{\mu \in p} \hat{S}_\mu^z$, and $1/2 < h < 1$.
To connect with the monomer-dimer model, we map the presence of a dimer on a link $\mu$ to $S^z_\mu = -1/2$ and the absence of a dimer to $S^z_\mu = +1/2$. 
Thus, the relation between the spins and the gauge fields is $2 \hat{S}_\mu^z = \hat{\sigma}^z_\mu$.
We assume $\Omega \ll V$ and treat $\hat{H}_1$ as a perturbation to the classical part $\hat{H}_0$.
Since there is a one-to-one correspondence between the sites of the Penrose tiling and the polygons, we use the terms ``Penrose sites" and ``polygons" interchangeably.
For a polygon $p$ hosting a monomer, all the spins on the vertices $\mu$ of $p$ will have $S^z_\mu = -1/2$, yielding $S_p^z = -q_p/2$. 
Thus, the contribution of such a polygon to the sum in $\hat{H}_0$ will be $V h^2$.
For a polygon $p$ touched by a dimer, $S_p^z = -q_p/2 + 1$, and its contribution to $\hat{H}_0$ will be $V (h-1)^2$.
Similarly, for a polygon touched by two dimers (an invalid configuration for a monomer-dimer model), its contribution to the sum in $\hat{H}_0 $ will be $V (h-2)^2$.
Since $1/2 < h < 1$, the energy of a site touched by a dimer is lower than that of a monomer, which is, in turn, lower than that of a site touched by two dimers.
Thus, it is energetically favorable for all sites to be touched by a dimer.
However, the geometry of the Penrose tiling prohibits perfect dimer coverings.
The next energetically favorable configuration is for a polygon to host a monomer.
Thus, the ground-state manifold of $\hat{H}_0$ is exponentially degenerate and consists of all maximal dimer coverings of the Penrose tiling.

Next, we consider the effects of quantum fluctuations of $\hat{H}_1$ on the effective Hamiltonian in the ground-state manifold.
At first order in $\Omega/V$, no new terms arise in the effective Hamiltonian because a single application of $\hat{S}_\mu^x$ on any maximal dimer configuration moves it out of the maximal dimer covering sector.
At second order in $\Omega/V$, the monomer hopping term shown in \cref{eq:dimer-monomer-H} is obtained.
At this order, the application of $\hat{S}^x$ twice on link $\mu$ generates a diagonal term in the effective Hamiltonian, corresponding to an energy shift of the maximal dimer configuration on which $\hat{S}^x$ is applied.
However, we find that this constant is not the same for all maximal dimer configurations, leading to a splitting of the degeneracy.
This term is undesirable if the goal is to obtain an effective Hamiltonian identical to $\hat{H}$ in \cref{eq:dimer-monomer-H}.
At third order in $\Omega/V$, no new terms are introduced.
At fourth order, we obtain both the ring-exchange term shown in \cref{eq:dimer-monomer-H} and undesirable dimer-configuration-dependent energy shifts.
To summarize, the classical ground state manifold consists of maximal dimer configurations, and the effective Hamiltonian in this manifold, up to fourth order in $\Omega/V$, consists of monomer hopping terms, ring-exchange terms, and dimer-configuration-dependent energy shifts.
We note that an alternative perturbation also generates an effective Hamiltonian with the terms mentioned above.
This alternative perturbation is
\begin{equation}
    \hat{H}_1' = \Omega' \sum_{\langle \mu, \nu \rangle} \hat{S}_\mu^+ \hat{S}_\nu^- + \hat{S}_\mu^- \hat{S}_\nu^+~,
\end{equation}
where the sum is over all pairs of links $\mu, \nu$ sharing a Penrose site.

While we do not obtain the exact Hamiltonian studied in the main text [\cref{eq:dimer-monomer-H}], this section demonstrates how a quantum monomer-dimer model on a Penrose tiling can be realized starting from spin degrees of freedom or qubits.
This is desirable because dimer degrees of freedom are not directly available in experimental platforms such as digital quantum computers, ion traps, optical lattices, Rydberg atom arrays in optical tweezers, etc.

The square-lattice quantum dimer model is known to undergo a phase transition as the monomer density varies~\cite{syljuasen2005continuous}. 
This motivates the study of quantum monomer-dimer models on Penrose tilings, where the density of mobile monomers---monomers that can hop---can be tuned. 
One way to achieve such a variable density of mobile monomers is by pinning monomers in maximal dimer coverings through site-dependent Zeeman fields added to the Hamiltonian in \cref{eq:microscopic-H}.
Specifically, if $\mathcal{M}$ denotes the set of Penrose tiling sites where we wish to pin the monomers in maximal dimer coverings, then the term to be added to the Hamiltonian of \cref{eq:microscopic-H} is:
\begin{equation}
    \hat{H}_{\text{pin}} = \alpha \sum_{i\in \mathcal{M}} \sum_{\mu \in \text{star}(i)} \hat{S}^z_\mu~,
\end{equation}
where $\text{star}(i)$ is the set of links with $i$ as one of their endpoints and $\alpha > 0$.
Notably, some sites---such as those in monomer membranes---can never host a pinned monomer in maximal dimer coverings.
Furthermore, in maximal dimer coverings, certain configurations of monomer locations are not allowed. 
For example, two neighboring sites can never both be occupied by monomers.
Thus, we require $\mathcal{M}$ to contain only sites where monomers can be pinned simultaneously in maximal dimer coverings.
In the limit $\alpha, V \gg \Omega$ ($V/\alpha$ can be large or small), the classical ground-state manifold consists of maximal dimer coverings with monomers pinned at sites in $\mathcal{M}$.
The sites where monomers are pinned can effectively be removed without affecting the dimer coverings.
More specifically, every maximal dimer covering of the Penrose tiling with monomers pinned on sites of $\mathcal{M}$ can be mapped to a maximal dimer covering of a modified Penrose tiling without any pinned monomers.
Here, by modified Penrose tiling, we mean the tiling obtained by removing the sites of $\mathcal{M}$ from the Penrose tiling and the edges that have an endpoint in $\mathcal{M}$.
The monomer-dimer model on the modified Penrose tiling will have a smaller density of monomers than the Penrose tiling.
Different choices of $\mathcal{M}$ thus enable control over the density of mobile monomers.

\end{document}